\documentclass[aps,twocolumn,amsfonts,amssymb,amsmath,showpacs,letterpaper,superscriptaddress,nofootinbib,altaffilletter,fontenc, floatfix, 10pt]{revtex4-2}

\usepackage{epsfig}
\usepackage{epstopdf}
\usepackage{graphicx}
\usepackage{orcidlink}
\usepackage{ulem}

\DeclareMathOperator{\E}{\mathbb{E}}

\begin{document}

\title{Parameter estimation from the core-bounce phase of rotating core collapse supernovae in real interferometer noise
}


\author{Laura O. Villegas} 
\email[E-mail: ]{laura.villegas8344@alumnos.udg.mx}

\author{Claudia Moreno} 
\email[E-mail: ]{claudia.moreno@academico.udg.mx}
\affiliation{Departamento de F\'isica, CUCEI, Universidad de Guadalajara
C.P. 44430, Guadalajara, Jal., M\'exico}

\author{Michael A. Pajkos} 
\email[E-mail: ]{mpajkos@caltech.edu}
\affiliation{TAPIR, Mailcode 350-17, California Institute of Technology, Pasadena, CA 91125}

\author{Michele Zanolin} 
\email[E-mail: ]{zanolinm@erau.edu}
\affiliation{Embry-Riddle Aeronautical University, Prescott, AZ 86301, USA}

\author{Javier~M.~Antelis}
\email[E-mail: ]{mauricio.antelis@tec.mx}
\affiliation{Tecnologico de Monterrey, Escuela de Ingeniería y Ciencias, Monterrey, N.L., 64849, México}


\date{\today}

%
\bigskip
\begin{abstract}
    In this work we propose an analytical model that reproduces the core-bounds phase of gravitational waves (GW) of Rapidly Rotating (RR) from Core Collapse Supernovae (CCSNe), as a function of three parameters, the arrival time $\tau$, the ratio of the kinetic and potential energy $\beta$ and a phenomenological parameter $\alpha$ related to rotation and equation of state (EOS). To validate the model we use 126 waveforms from the Richers catalog \cite{Richers_2017} selected with the criteria of exploring a range of rotation profiles, and involving EOS. To quantify the degree of accuracy of the proposed model, with a particular focus on the rotation parameter $\beta$, we show that the average Fitting Factor (FF) between the simulated waveforms with the templates is 94.4\%.  In order to estimate the parameters we propose a frequentist matched filtering approach in real interferometric noise which does not require assigning any priors. We use the Matched Filter (MF) technique, where we inject a bank of templates considering simulated colored Gaussian noise and the real noise of O3L1. 
    For example for A300w6.00\_BHBLP at 10Kpc we obtain a standar deviation of $\sigma = 3.34\times 10^{-3}$ for simulated colored Gaussian noise and $\sigma= 1.46\times 10^{-2}$ for real noise.
    On the other hand, from the asymptotic expansion of the variance we obtain the theoretical minimum error for $\hat{\beta}$ at 10 kpc and optimal orientation. The estimation error in this case is from $10^{-2}$ to $10^{-3}$ as $\beta$ increases. We show that the results of the estimation error of $\beta$ for the 3-parameter space (3D) is consistent with  the single-parameter space (1D), which allows us to conclude that $\beta$ is decoupled from the others two parameters.
\end{abstract}


\maketitle

\section{Introduction}
\label{Sec:Introduction}

The LIGO/Virgo/KAGRA collaborations \cite{TheLIGOScientific:2014jea, TheVirgo:2014hva, Aso:2013eba} have detected multiple Gravitational Waves (GW) events emitted by binary neutron stars (BNS), binary black holes (BBH) and neutron star black hole binaries (BNSBH) \cite{LIGOScientific:2021qlt, LIGOScientific:2021psn, https://doi.org/10.48550/arxiv.1811.12907, Abbott_2021, https://doi.org/10.48550/arxiv.2111.03606}. Research in multimessenger physics including GWs commenced following the observations of BNS and BNSBH \cite{Abbott_2016}. The search for GW generated by other exotic astrophysical sources, such as gamma-ray bursts \cite{Urrutia:2022lce} and Core Collapse Supernovae (CCSNe) \cite{ Abdikamalov_2014, Srivastava_2019, JANKA_2007, Kuroda_2016} is ongoing and will continue in the next observing runs. CCSNe are expected to expand the multi-messenger astronomy program \cite{Szczepa_czyk_2021}. CCSNe are highly energetic explosions generated by a progenitor star with a mass greater than $\sim 8 M_\odot$ \cite{Scheidegger_2010, O_Connor_2018, Couch_2020, Warren_2020}. Although promising engines are proposed to drive these explosions (for example, the delayed neutrino heating mechanism \cite{Bethe_1985} and the magnetorotational mechanism \cite{Bisnovatyi-Kogan_1970,leblanc_1970}), the finer details of these explosion mechanisms are not yet fully understood \cite{Radice_2016}. Some physical phenomena involved in CCSNe include protoneutron star (PNS) convection, prompt convection in the post-shocked material, neutrino-driven convection later on ($\gtrsim 50$ ms after bounce) \cite{burrows:1992},  g-mode oscillations, the standing accretion shock instability (SASI), rotational flattening of the bouncing core, and non-axisymmetric rotational instabilities \cite{Foglizzo_2015, Radice_2019}. 

Several groups have performed simulations of CCSNe with slowly rotating progenitors \cite{Couch_2014, Fuller_2015, andresen_2018, Boccioli_2023}, as well as with Rapidly Rotating (RR) progenitors \cite{Dimmelmeier_2008, Dimmelmeier_2002,Ott_2012, powell_2016,gilkis_2018, kotake_2020, harada_2019, harada_2022, wang_2024, Zwerger:1997sq}. Although slowly rotating progenitor simulations have been performed for time scales of hundreds of milliseconds, the simulations for RR CCSNe progenitors were initially performed for the first few tens of milliseconds to capture the GW bounce signal and more recently extended to longer time scales \cite{takiwaki_2018, pajkos_2019, pan_2021, takiwaki_2021, hsieh_2024, PhysRevD.103.024025, PhysRevD.93.042002}, some with the inclusion of magneto-hydrodynamics (MHD) routines \cite{mosta_2014, kuroda_2020,matsumoto_2023,nakamura_2024,obergaulinger_2020, powell_2020, powell_2023, bugli_2023}.

During the supernova accretion phase, the reconstructed GW emission primarily exhibits a stochastic nature in the time domain. However, deterministic features in the time domain arise for rotating models just after bounce and are apparent in the frequency domain for rotating and non-rotating models \cite{mezzacappa2024gravitational}. One particularly important feature, that is expected to be drastically different in RR CCSNe versus slowly rotating ones, is the core bounce component occurring in the first $\lesssim 10$ milliseconds, which may have a large amplitude due to the deformation of a rotationally distorted PNS interacting with infalling material. 
CCSNe simulations have revealed that the gravitational signal from a RR CCSNe has a strong and systematic dependence on the degree of differential rotation, described by the core-bounce dynamics, and that it is related to the ratio of rotational kinetic energy to gravitational energy $\beta=T/|W|$ \cite{Abdikamalov_2014, Richers_2017, Shibagaki_2020}, known as the rotation parameter.

According to \cite{Abdikamalov_2014}, for RR progenitors, due to the axisymmetric geometry of the initial perturbation, the PNS pulsations remain axisymmetric in the core-bounce phase, and 2D axisymmetric simulations can accurately predict the GW signal in this phase.

The physical parameter estimation analysis during the core-bounce phase of RR CCSNe aims to estimate information about the degree of rotation in the progenitor and about the PNS Equation of State (EOS). An important part of the process is to choose templates that match the GW signal embedded in realistic interferometric noise. These templates could be obtained directly from numerical simulations (the simulation reliability may have slight changes over time as the level of sophistication increases) or analytical templates, depending on phenomenological parameters. 

In ref. \cite{Chao:2022tui}, authors used waveforms from the Richers catalogue \cite{Richers_2017} and Convolutional Neural Network (CNN) methods without any noise to identify fundamental rotational characteristics, defined by 2D rotation profile, maximum initial rotation rate ($\Omega_0)$, the degree of differential rotation (A), and the nuclear EOS. They explore gravitational wave signals using two time periods $(-10, 6)$ ms and $(-10, 54)$ ms. The first correspond to core-bounce, and the second one is an additional $48$ ms, to include the prompt convection phase. Using this CNN method, their results indicate that the accuracy of calculated values suggests the GW signal from the bounce can diagnose the core angular momentum of the collapsing progenitor and is minimally affected by the nuclear EOS. 

In a recent arXiv \cite{pastormarcos2024bayesian}, the parameter estimation is carried out considering that the waveform is embedded in additive Gaussian noise. The analytical model presented has a linear dependence between the difference of the amplitude of core-bounce peaks $D\,\Delta h$ and $\beta= T/|W|$, this corresponds to a sufficiently slow rotation ($T/| W| < 0.06$). The waveforms in the initial phase post-bound are characterized using an analytical function that depends on two parameters  $D\Delta h$ and $f_\mathrm{peak}$. 
The analytical expression presented in that work coincides with the core-bounce phase and the oscillations during the post bounce.
Through a Bayesian inference analysis using the informative priors, but no studies of the effect of varying the priors on the final results, they estimate the parameters $f_\mathrm{peak}$, $D\Delta h \, \sin^2\theta$, and $\Psi$ with a precision better than 10\% for more than 50\% of the  considered events at distances in the range $0.1-1000$ Kpc.

In this work, we operate in the presence of real LIGO noise. We limit the analysis to the deterministic part of the waveform, that is, only the core-bounce phase. As mentioned, it is closely related to the parameter $\beta$ but also depends on the rotational profile as well as the EOS. We include waveforms with a different degree of rotation which includes slow and rapid rotation regimes, defined in \cite{Richers_2017}. While stellar evolution calculations highlight the importance of magnetic breaking to slow the internal rotation of single stars \cite{woosley:2006}, the influence of binarity on stellar evolution, and its influence on stellar rotation remains an ongoing area of research \cite{sana_2012, song_2016}.  To account for this wide array of uncertainty in CCSNe initial conditions, we consider a wide range of possible $\beta$ ($0.02 < \beta < 0.14$). 
We introduce a phenomenological parameter $\alpha$ that describes the dependence in the third core-bounce peak on that immediately precedes post-bounce oscillations. This parameter $\alpha$ allows us to analyze beyond the linear dependence between the parameters $D\,\Delta h$ with $\beta= T/|W|$.  
The next CCSNe close enough to be detected will have enormous scientific relevance \cite{Roma_2019, powell_2016}. If the reconstruction of the signal were sufficient to establish whether the parameter $\beta$ is different from zero, such an observation would allow establishing the presence of rotation in the progenitor, providing a valuable constraint on the role of rotation in stellar evolution modeling.

We use a frequentist approach which does not require priors (that cannot be derived from previous observations GW CCSNe since we do not have them yet). 
We use quadratic functions that adjust to the phenomenological characteristics of the waveform peak. The spirit of this paper is to model the core bounce phase waveform with the simplest analytical model capturing the essential features of the core bounce and in terms of a three dimensional parametrization. 
We investigate the error in estimating $\beta$, as well as other parameters characterizing the core bounce GW component, in two directions: first, the performance in colored and real interferometric noise, of a classical matched filtering approach \cite{doi:10.1142/S0218271823400060}. Second, we show the theoretical minimum error regardless of the algorithm that someone might employ \cite{Vitale_2010}. Historically, the main tool adopted for the second task is called the frequentist Cramer Rao Lower Bound (CRLB) and predicts the minimum variance of any unbiased estimator. It can, however, underestimate the real error for non-linear estimation or non-Gaussian data. A technique that can complement the CRLB is the usage of asymptotic expansions; we adopt it here. Explicitly, it involves the use of asymptotic expansions for errors of the Maximum Likelihood Estimator (MLE). This methodology has previously been applied to the GWs from binary black holes \cite{Vitale_2010, Vitale2011} to quantify the precision in estimating physical parameters such as total mass, reduced mass, and phase \cite{Zanolin_2010}. The use of asymptotic expansions allows one not only to predict the minimum possible errors in the estimation of these parameters as a function of the noise spectra of the interferometers and the distance from the source, but also to evaluate how far the estimation algorithms are from reaching the lower limits of error. When parameter estimation is conducted using matched filtering, with signals accurately represented by analytical templates and influenced by additive noise, this method is an optimal parameter estimation approach and is also equivalent to a MLE. Frequentist parameter estimation is also used for new discoveriesn in particle physics (see, for example, \cite{Feldman_1998}) and in supernova searches with coherent Wave Burst (cWB) who is frequentist as well \cite{Necula_2012, DRAGO2021100678}.

There are indications that inference of rotational properties from the rate of increase in PNS resonant frequencies will require understanding the role of EOS as well as progenitor properties from more than one GW feature \cite{Markakis_2009, casallas2023}. Because of this, exploring the inference potential from different signal features and different multimessengers, such as neutrinos, is important, also given the different role of noise in all GW and neutrino channels. 

The structure of the paper is the following. In Section \ref{Sec:MLEerrorsCCSNe} we analyze 126 waveforms selected from 1824 simulations listed in the Richers catalog \cite{Richers_2017}. These models are selected for their EOS, which align with both observational data from neutron stars and constraints from nuclear physics experiments. Further details on this selection process are provided in the Appendix \ref{appendix:waveform}.

In Section \ref{Sub:model}, we introduce an analytical model designed to fit the core-bounce of the chosen numerical waveforms using three phenomenological parameters. These templates enable the computation of the first order variance using the CRLB and the second order variance through the derivatives of the aforementioned analytical function. These templates are also used to perform the frequentist MF parameter estimation. 
In Section \ref{Sub:FF}, we quantify the degree of ambiguity of the CCSNe signals chosen in this work, similar to studies performed with template banks for coalescing compact binaries. In this paper, we quantify this in two ways: we calculate the FF \cite{PhysRevD.52.605, Cho_2018} for the value of the parameters that best fit the numerical waveform (versus the nominal ones adopted in the numerical simulation).  For one of the parameters $\beta$, we produce the histogram of the true versus MF estimate. 
In Section \ref{Sub:MF}, we use matched filtering to estimate $\hat{\beta}$ parameter at different distances. Finally, in Section \ref{SubSec:Parameter} we calculate the estimation error for the rotation parameter $\beta$, applying the asymptotic variance expansion. 
In Section \ref{Sec:AsymptoticExpansions}, we present the Cramer Rao Lower Bound approach, and the asymptotic expansion corrections when we have a signal in colored noise defined by the Power Spectral Density (PSD) of the detector. 
Our conclusions are presented in Section \ref{sec:conclusions}.

\section{Core Bounce component of the GW from a rapidly rotating CCSNe}
\label{Sec:MLEerrorsCCSNe}
 The GW signature of RR CCSNe has a rather simple deterministic morphology in the first few milliseconds after bounce. It is characterized by three consecutive extrema around the core bounce phase, an increase in amplitude before the core bounce, a large depletion during the bounce, and an increase in amplitude as the core rebounds. The temporal separation of the extrema has a negligible variability between waveforms as discussed in Appendix B.
 Subsequently, the energy of the core is dissipated hydrodynamically, which is characterized by pulsations known as ring-down. During the ring-down, the signal is more stochastic and, therefore, more difficult to describe analytically.

\begin{figure}[!ht]
 \centering
 \includegraphics[width=8.0cm]{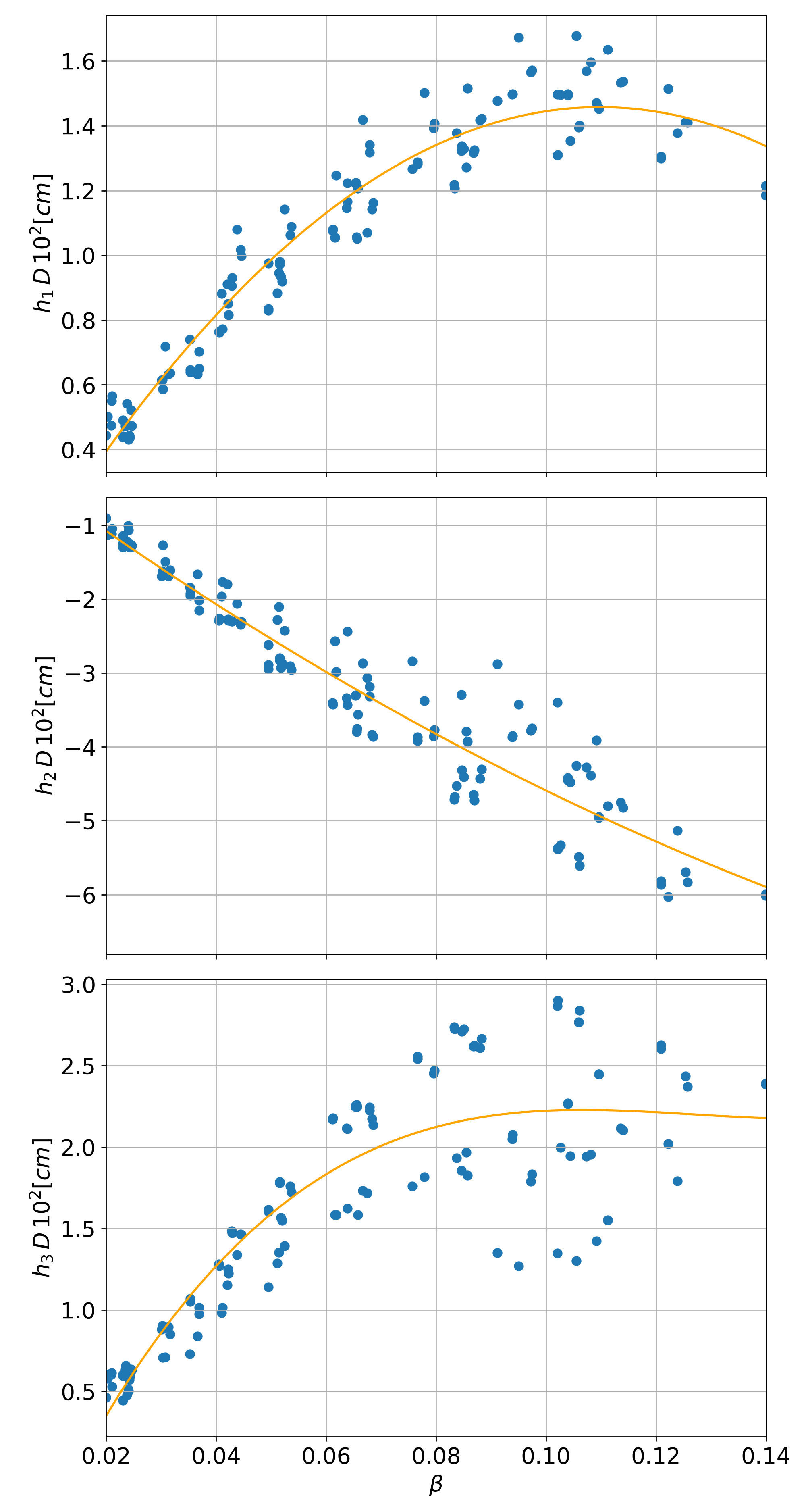}
 \caption{Relationship between the parameter $\beta$ and the amplitude peaks of the core bounce. The three images represent the amplitude of the three peaks of the GW strain; the dots are the values obtained in the 126 simulations, and the solid line is the quadratic function that fits the behavior of each peak. 
 }
\label{Fig:betavshp}
\end{figure}

One of the important characteristics of the waveforms is how the amplitude of each peak in the core-bounce relates to the rotation parameter $\beta$, as shown in Figure \ref{Fig:betavshp} and discussed in \cite{Abdikamalov:2020jzn}. This parameter $\beta$ captures how deformed the PNS is due to rotation, while quantifying the restoring force of gravity, which ultimately dictates the dynamics of the core bounce. The natural question arises: 
Can angular velocity also be considered a deterministic parameter? While the angular velocity profile can provide an intuitive sense of rotation, it must be tied to the mass distribution within the supernovae to capture the strength of the gravitational restoring force during the rebound. Therefore, while PNS spin periods (via angular velocities) can be estimated from assumed PNS moments of inertia \cite{Guilet_2014} or physics-based simulations \cite{Rantsiou_2011, Kazeroni_2016}, $ \beta$ still is a useful quantity because it contains information about radial profiles of mass density and rotational speed.

Stiffer equations of state are expected to support larger PNS radii for the same mass \cite{Lattimer_2012}. Consistently, we expect that the dynamics of a rotating CCSNe to be affected
by the stiffness of the EOS. Figure \ref{Fig:betavshp} indicates that for low values of $\beta$ a one dimensional parameter space could be sufficient to predict the peak amplitudes.  However, for larger $\beta$  a larger parameter space is needed. In cases when $\beta \gtrsim 0.1$, the core becomes more centrifugally supported at bounce and the GW bounce signal depends much more strongly on the amount of precollapse differential rotation.
A proposal is made in the next section to model this signal.

%

\subsection{Core-Bounce signal analytical model}
\label{Sub:model}
We used data from 126 numerical simulations of GW generated by \cite{Richers_2017}, which were chosen considering six EOS and 16 rotation profiles.  

We chose those that most closely align with observational constraints of neutron stars and experimental constraints on nuclear EOS to build a parameter space that characterizes the core-bounce phase for the case of rapidly rotating CCSNe.  For more details on our selection, see Appendix \ref{appendix:waveform}.

The Richers catalogo consider a progenitor with $12 M_\odot$. The core-bounce components only depends of amount of angular momentum in the core with needs a $0.4 M_\odot$ at the center of star, so this database is valid in general.

In principle, one could suggest that the parameter space of the waveforms is six dimensional with three amplitudes and three times. However, in terms of morphology, the first arrival time is not relevant, and the delays of the second and third peak with respect to the first are more relevant. This makes it five.

\begin{figure}[!ht]
 \centering
 \includegraphics[width=8.0cm]{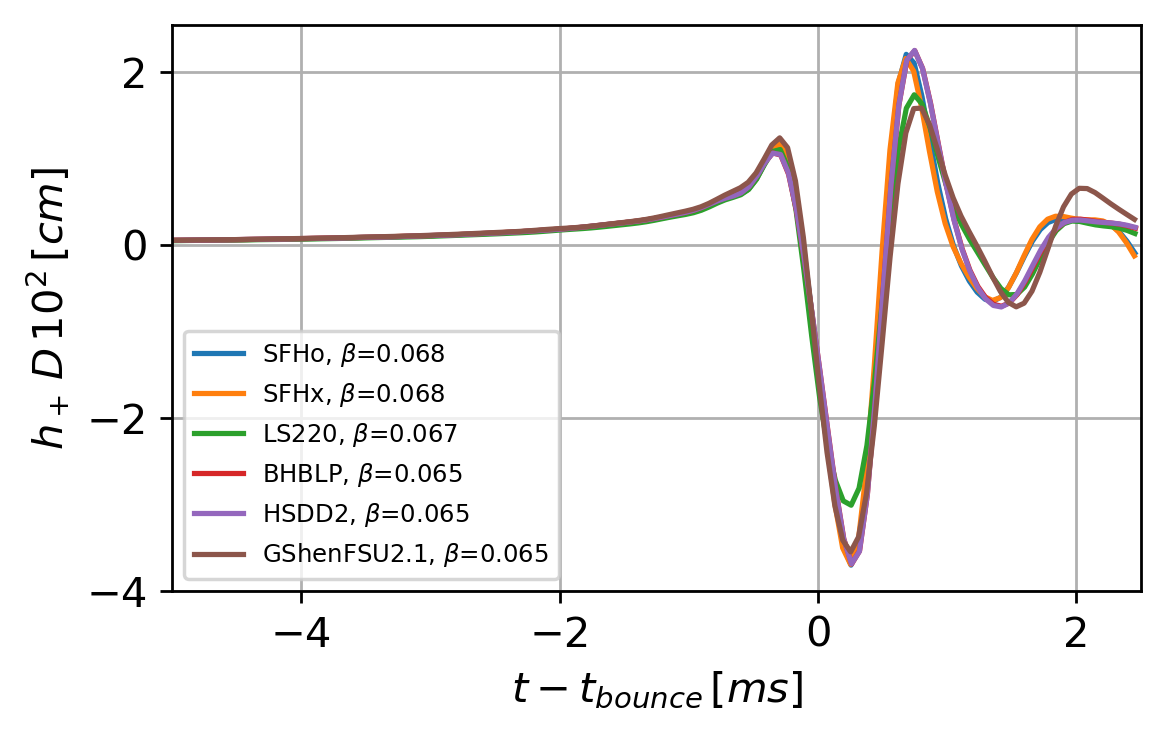}
 \caption{A300w6.00 waveforms taken from the catalog of Richers for different EOS. The variation that exists in the second and third peaks are due to EOS.  
 }
\label{Fig:differentEOS}
\end{figure}

\begin{figure}[!ht]
 \centering
 \includegraphics[width=8.0cm]{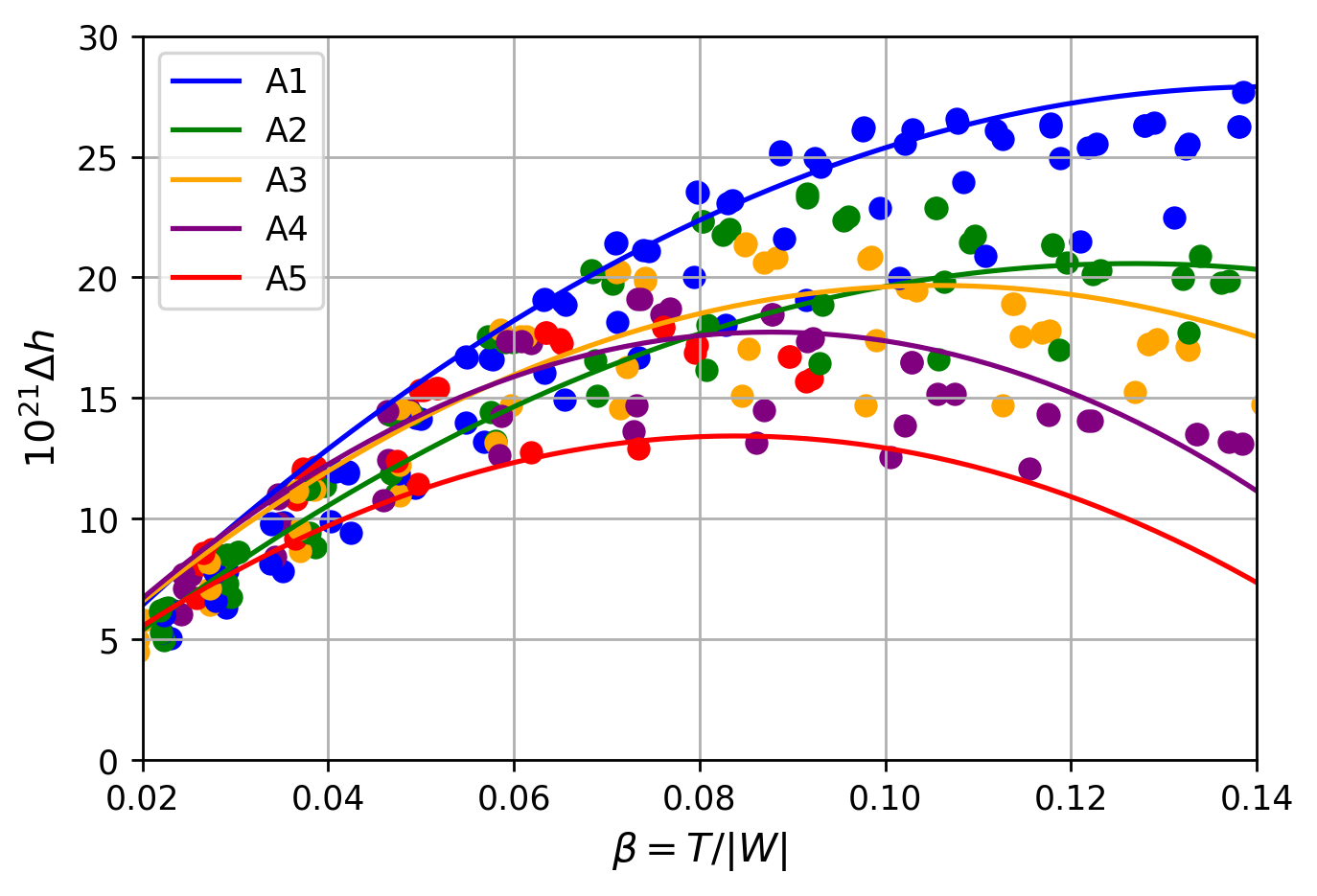}
 \caption{Relationship between the parameter $\beta$ and $\Delta h(t)$ bounce, which corresponds to the horizontal difference that exists between the first $h_1(t)$ and last peak $h_3(t)$. For each rotation profile $A_n$, $n=1,..,5$, we consider the six different EOS. }
\label{Fig:betavsdeltah}
\end{figure}

If the three peaks are equally spaced, it becomes four dimensional. Furthermore, the amplitude of the peaks are not strictly independent.  As an example, consider Figure \ref{Fig:differentEOS}, which shows sample bounce signals for different EOS. 
 When scaled by distance, the first three extrema of the signal occur around $h_1 \sim 100$ cm, $h_2 \sim -350$ cm, and $h_3 \sim 200$ cm.  By taking the maximum between $h_1$ and $h_3$, then subtracting $h_2$, one obtains a different, observable metric $\Delta h$, commonly used in the study of GW bounce signals.  More concretely, $\Delta h = \max(h_1, h_3) - h_2$.  As supported by \cite{Richers_2017}, and seen in Figure \ref{Fig:betavsdeltah}, $\Delta h$ can be approximated by a linear fit. As the amplitudes of the three peaks are not independent, the residual possible dimensionality goes down from four to three. In this paper we test the goodness of this three dimensional parameter space with the calculation of the FF, the same metric adopted in Compact Binary Coalescence (CBC) searches to assess the goodness of the template bank \cite{Roulet_2019}. 
Given the expected challenges of data analysis for CCSNe in real interferometric data,
the goal of this analysis is to identify the smallest parameter space that
describes the GW and establish if in real non-Gaussian noise we can estimate the parameters describing the degree of rotation and possibly properties related to the EOS with a few tens of relative errors. 

We propose an analytical model that categorizes the bounce phase of the PNS in terms of three parameters. Looking for a simple and differentiable function that can model the core-bounce, Gaussian functions are chosen, where we can control the amplitude of the characteristic peaks, the time to occur, the duration of the signal and the relationship of the signals with the different EOS, represented by
\begin{eqnarray} \label{Ec:signalmodel}
 h(t) &= h_1(\beta)\,{\rm exp}^{\left[-\frac{(t - \tau)^2}{2\eta^2} \right]} + h_2\,(\beta)\,{\rm exp}^{\left[-\frac{(t - \tau_a)^2}{2\eta^2} \right]} \nonumber \\ 
 & + h_3(\alpha, \beta)\,{\rm exp}^{\left[-\frac{(t - \tau_b)^2}{2\eta^2} \right]}\,,
\end{eqnarray}
where $\eta = 0.2$ ms and $\tau$ correspond to the horizontal displacement of the first peak. We consider that this first peak occurs when $\tau$ is between -0.5 to -0.2 milliseconds. The position of the second and third peak can be obtained in terms of $\tau$, so that we define $\tau_a = \tau + 0.5 $ and $\tau_b= \tau + 1$, for $h_1(\beta) $ and $ h_2(\beta)$. The fixed temporal separation of the peaks is supported by the spectral analysis discussed in appendix \ref{Appendix:freq}. The factor $h_3(\alpha, \beta)$ describes the amplitude of the third peak.  Figure \ref{Fig:betavshp} shows that the three peaks depend only on $\beta$ in the low $\beta$ limit but a larger dimensional space is needed for larger $\beta$, so we will perform an extra analysis that allows us to define a new phenomenological parameter $\alpha$, which is linked to the change in amplitude due to different EOS. We explain the details of $h_3(\alpha, \beta)$ later.   

To obtain the amplitude functions of each peak $h_i(\beta)$, $i=1,2$, we use the chosen simulated waveforms in \cite{Richers_2017} and relate the amplitude peaks to the $\beta$ values, Figure \ref{Fig:betavshp}. We then fit an analytical curve to the data.  This method allows us to have two second order polynomials  which  adequately describe the behavior of the amplitude as a function of the rotation parameter. Since we will use higher than second order derivatives to calculate the covariance, we choose differentiable functions that allows the mathematical calculations. To describe the amplitude of the two first  peaks in our analytical model, we use fit a quadratic function as,
\begin{eqnarray} \label{Ec:Amplitudes}
    h_1(\beta) &=& -13.2 +  2.89\times 10^3 \beta -1.31\times 10^4 \beta^2 \,,  \\
    h_2(\beta) &=& -1.03 - 5.52\times 10^3 \beta + 9.43\times 10^3 \beta^2 \,. \nonumber 
\end{eqnarray}
However, when analyzing the waveforms in the catalog, we observed that there is a significant difference in the amplitude in the third peak. As an example, we used the A300w6 signal with different EOS, Figure \ref{Fig:differentEOS}. 
With this we want to relate the peak amplitudes with the third peak; for this we use the difference between the highest and lowest points in the bounce signal strain, as a function of parameter $\beta$.

In Figure \ref{Fig:betavsdeltah} we show the difference between the third peak and the first one in the core-bounce signal $\Delta h$ for the waveforms used in this work. We can do a polynomial interpolation for each rotation profile $A_n$, $n=1,..,5$ \cite{Richers_2017}, and define a curve in terms of $\beta$. But also we noticed that the EOS changes the function, since it generates greater dispersion. To account for this dispersion, we fit a quadratic function in $\beta$ with an additional variable $\alpha$ for the amplitude of the third peak (see Equation \eqref{Ecu:h3}). 

\begin{figure}[!ht]
 \centering
 \includegraphics[width=8.0cm]{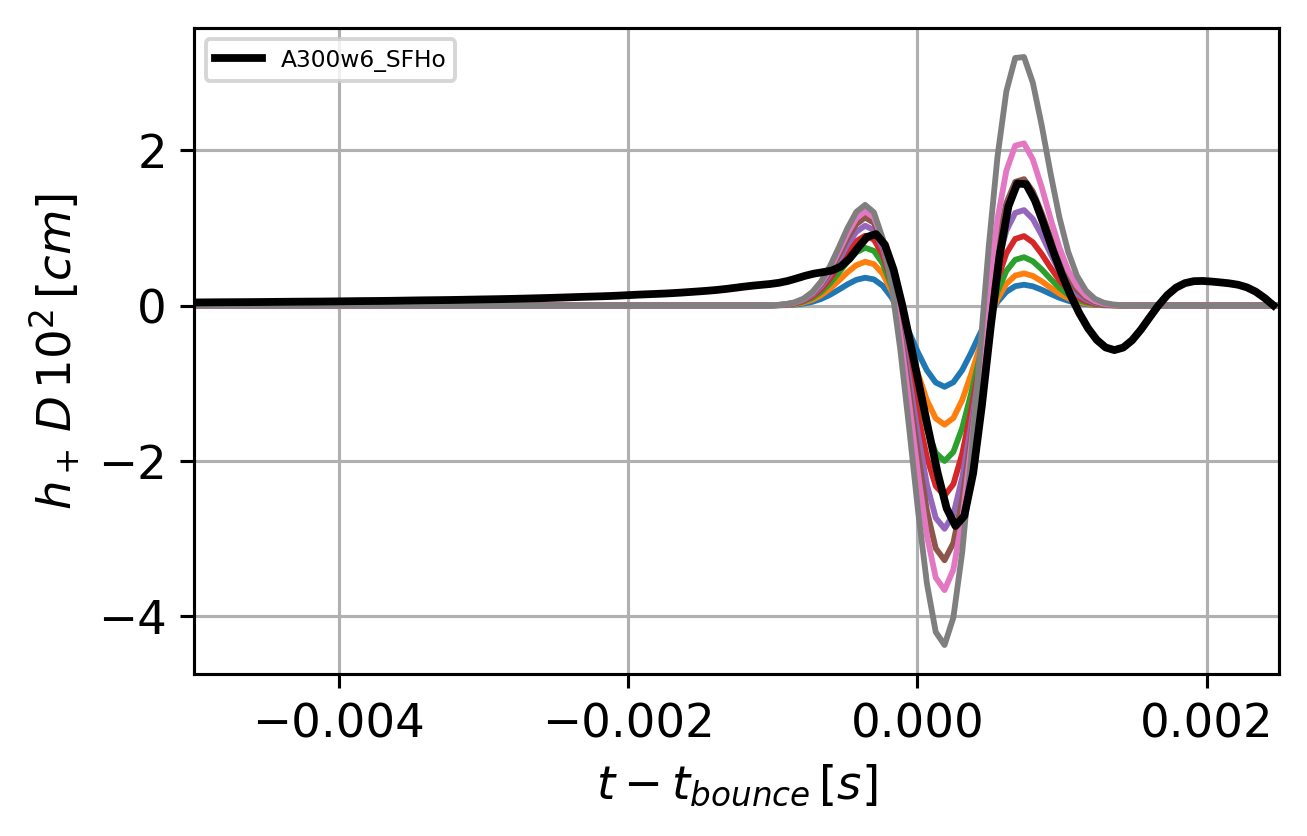}
 \caption{Waveform A634w6.00\_SFHo superimposed on a set of waveforms modeled with the analytical function at different values of the parameter $\beta$ (colored lines) with $\alpha$ and $\tau$ fixed.  We observe that analytical model generates similar morphology to the simulation waveform for the core-bounce phase (black line).}
\label{Fig:analyticalmodel}
\end{figure}

We define the function in terms of $\alpha$ and $\beta$, such that we have
\begin{equation}
    h_3(\alpha, \beta) =  17.20 + \alpha \left(\frac{\beta}{0.06} \right)^2 \,.
    \label{Ecu:h3}
\end{equation}
To summarize our new range of parameters, we have newly calculated values of $\alpha$ in a range $[30, 380]$, $\beta$ within $[0.02,0.14]$, and finally $\tau$ between $[-0.5, -0.2]$ms. In Figure \ref{Fig:analyticalmodel} we observe the simulated waveform A634w6.00\_SFHo superimposed on a set of waveforms modeled by Equation \eqref{Ec:signalmodel} we see that the analytical model reproduce the morphology of the core-bounce when we use a combination of the parameter values.

In the next section, to quantify the reliability of the model, we calculate the FF for the parameter $\beta$. In addition, we use matched filtering to estimate the parameter values. 

\subsection{Fitting Factor (FF)}
\label{Sub:FF}
\begin{figure}[!ht]
 \centering
 \includegraphics[width=8.0cm]{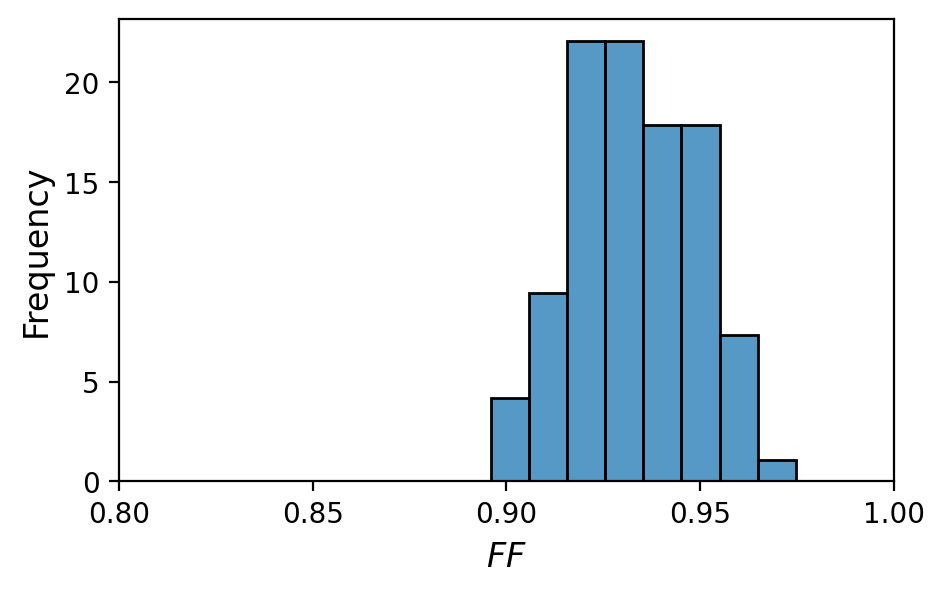}
 \includegraphics[width=8.0cm]{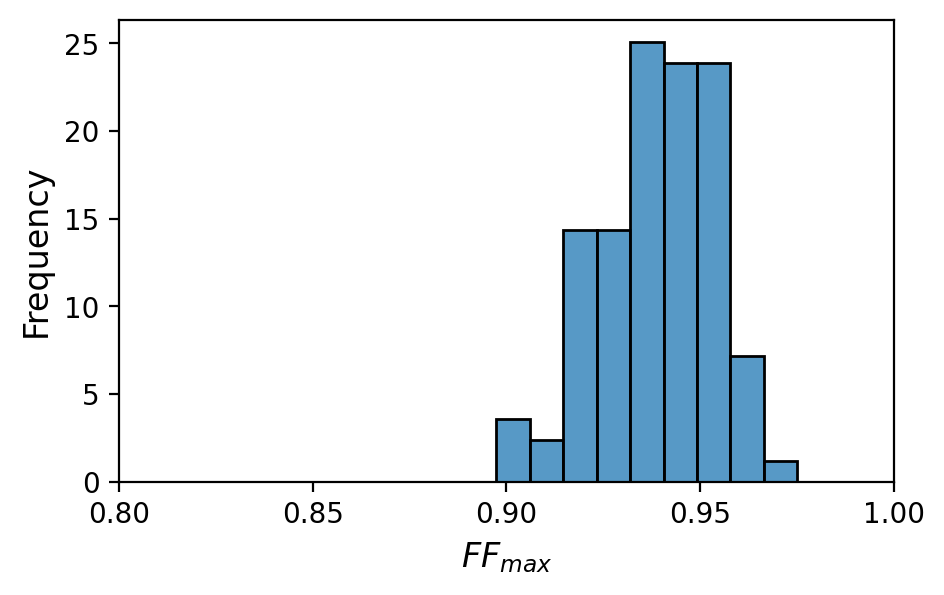}
 \includegraphics[width=8.0cm]{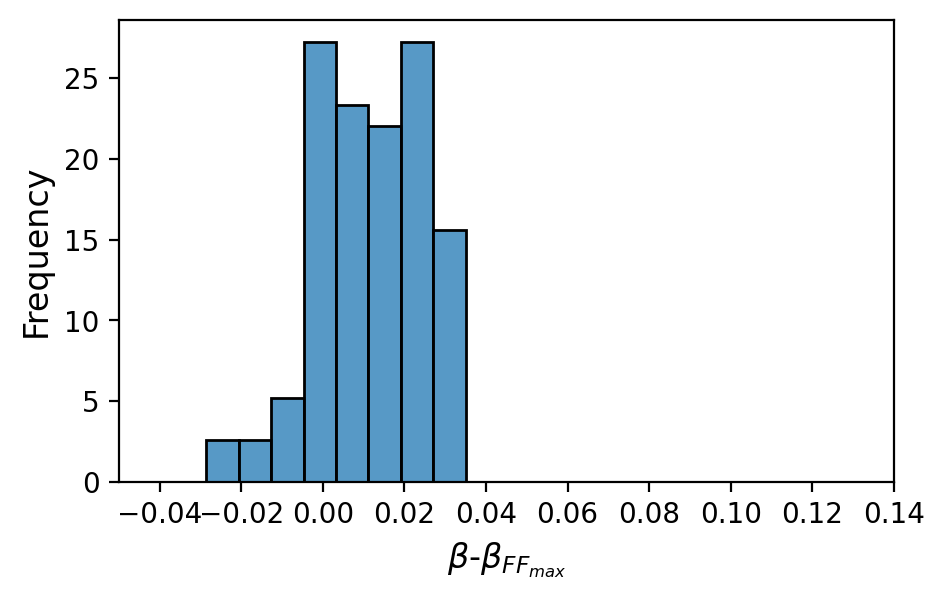}
 \caption{Top panel: Histogram of the values of the FF that compares the waveform generated by model with the real  parameter value and the simulation waveform. The average FF is 94.4\%. Middle panel: Histogram of maximum ${\rm FF}_{max}$. Where we considered the best combination of parameters in our model that fit each catalog's waveforms. Bottom panel: Histogram characterizing the degree of ambiguity of the template bank with FF. We obtain it from the subtraction between the real value $\hat{\beta}$ of and the estimated value of $\hat{\beta}$.
 }
\label{Fig:FF}
\end{figure}

The FF is a technique which calculates the similarity between two signals. In our context, we consider the frequency domain waveform obtained from the numerical simulations $h_s(f)$ and the corresponding analytical model proposed in this work $h_m(f)$. The FF is defined as
\begin{equation}
    \mathrm{FF} = \frac{\langle h_s(f)|h_m(s) \rangle}{\sqrt{\langle h_s(f)|h_s(s) \rangle\langle h_m(f)|h_m(s) \rangle}} \,,
\end{equation}
where FF=1  implies that the two signals are equal and we would have 100\% similarity.
We  first compute the FF between eachof the simulated signals from Richers catalog and its equivalent signal model using the simulations parameters $\alpha$, $\beta$ and $\tau$. The distribution of FF values is shown in Top panel of Figure \ref{Fig:FF} where the average FF is 94.4\%.

Another metric we produce is the best combination of parameters.  For this, we generates a group of templates using the parameter ranges defined above in \ref{Sub:model}, and we compare it with a specific waveform of Richers catalogue in such a way that we take the one where the FF is maximum $FF_{\rm max}$; it means the best combination of three parameters that match with the real waveform. We repeat this process with each of the 126 selected waveforms.  We find the maximum FF for the simulated waveforms and the template bank with the best combination of parameters, Figure \ref{Fig:FF} in middle panel. We obtain an average $FF_{\rm max}$ of 94.3\%. 

To assess the uncertainty in our analytical model regarding the $\beta$ parameter, we illustrate the distribution of the actual $\beta$ values subtracted by the optimal estimated rotation parameter $\hat{\beta}$ values for each signal, as shown in Figure \ref{Fig:FF}in bottom panel. 
%
\begin{figure}[th] 
 \centering
 \includegraphics[width=8.0cm]{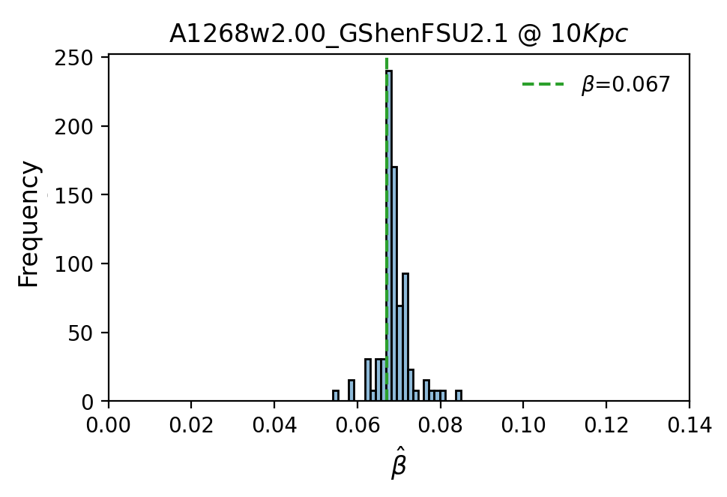}
 \includegraphics[width=8.0cm]{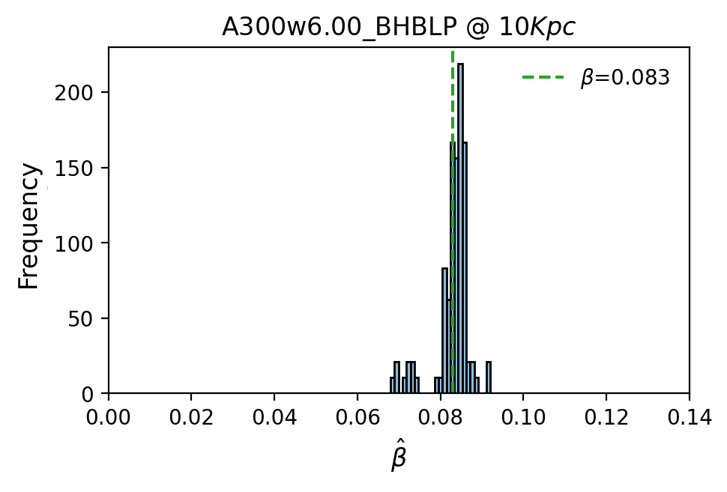}
 \includegraphics[width=8.0cm]{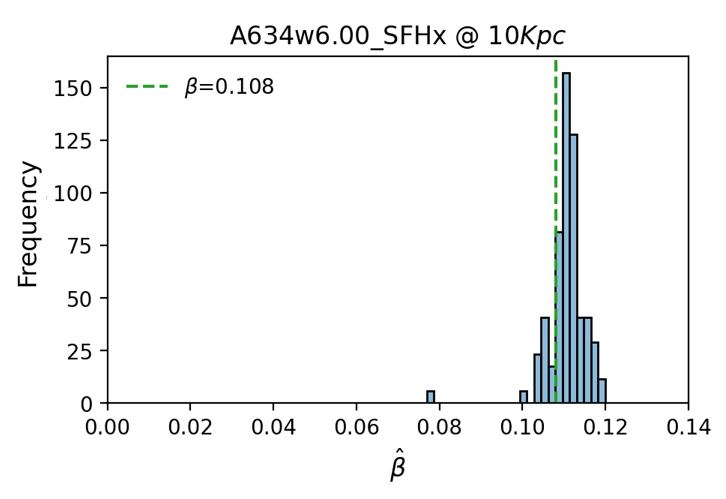}
 \caption{Histogram of the estimated values of $\beta$ with MF using colored simulated Gaussian noise. Vertical dotted lines represents the real $\beta$ values. Top panel: For waveform A1268w2.00\_GShenFSU2.1 we have an estimation average value of $\hat{\beta}=0.069$ and $\sigma = 5.39\times10^{-3}$. The middle panel: Correspond to waveform A300w6.00\_BHBLP with an estimation average value $\hat{\beta} =0.083$ and $\sigma = 3.43\times10^{-3}$. Bottom panel: Finally, for waveform A634w6.00\_SFHx the estimation average value is $\hat{\beta}=0.111$ with $\sigma = 2.01\times10^{-3}$.}
 \label{Fig:MF_colorednoise}
\end{figure}
%

\section{Matched Filtering Parameter Estimation}
\label{Sub:MF}

MF analysis is a basic tool for detecting known waveforms from a noise-contaminated signal and estimating its physical parameters \cite{Vitale_2010, Zanolin_2010, Vitale2011}. 

Using MF to estimate the value of $\hat{\beta}$, we utilize the waveforms selected in this study, which are injected into two cases: 1) colored simulated Gaussian noise using the PSD for O3 and 2) real LIGO O3L1 noise  

With our analytical model, we build a bank of 90000 templates at different distances. Then, we find the best template that fits the simulated signal. To illustrate the outcomes derived from estimating the $\alpha$ and $\beta$ parameters, we showcase three different scenarios: the first is shown in the top panel of Figure \ref{Fig:MF_colorednoise} which corresponds to the distribution of the estimated parameter $\hat{\beta}$ taking the simulated waveform A1268w2.00\_GShenFSU2.1 at a distance of 10 kpc and a value of $\beta=0.067$.  That is, it is in the slow rotation regime $\beta<0.08$, according to \cite{Richers_2017}. The second example is shown in the middle panel of Figure \ref{Fig:MF_colorednoise}, where the chosen waveform is A300w6.00\_BHBLP at a distance of 10 kpc with $\beta=0.083$. Finally, we take the waveform A634w6.00\_SFHx at a distance of 10 kpc and a value of $\beta=0.108$, bottom panel in Figure \ref{Fig:MF_colorednoise}. The last two examples are in the rapidly rotating regime for $0.08<\beta<0.12$ \cite{Richers_2017}. The histograms show the distribution of estimated values $\hat{\beta}$ obtained from the simulations and the real value $\beta$ (green line). 

We observe that the distribution of estimated values is close to the mean value of the estimates. Quantitatively, the standard deviation in the case of colored simulated Gaussian noise is of the order of $10^{-3}$.

\begin{figure}[!ht]
 \centering
 \includegraphics[width=8.0cm]{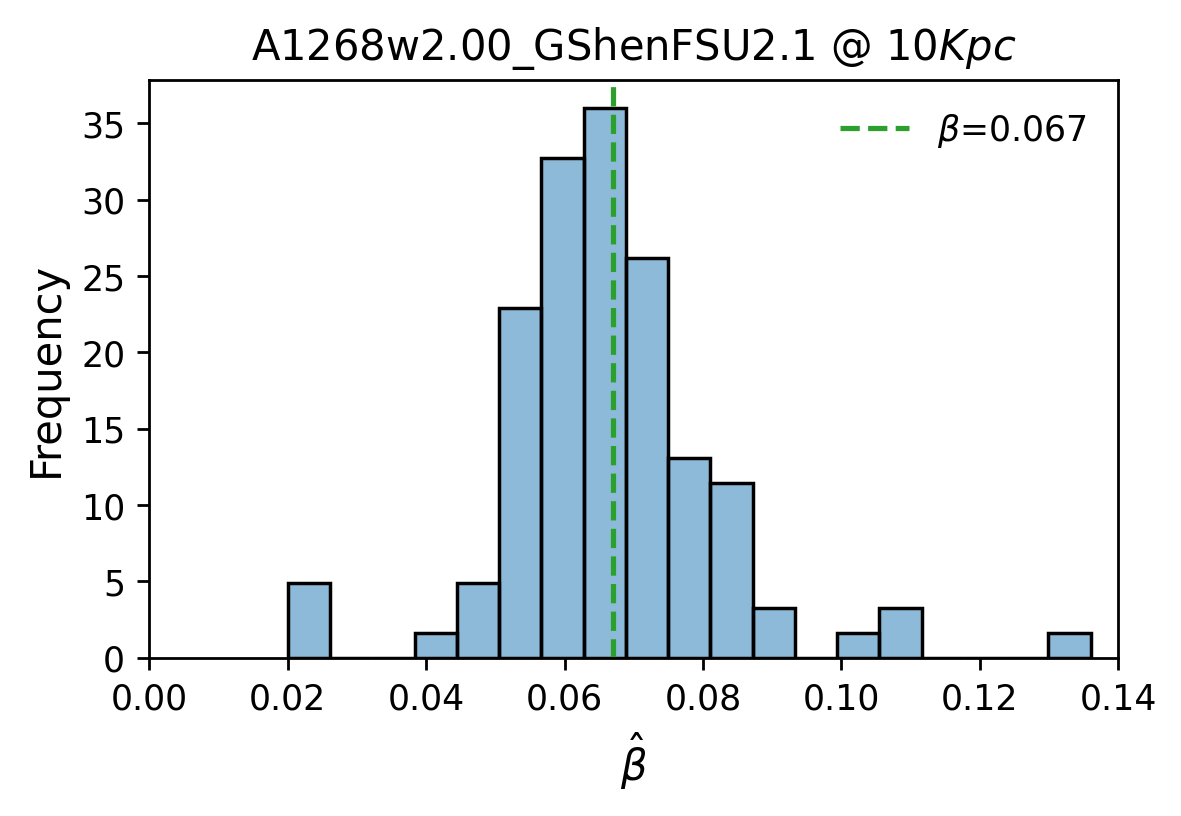}
 \includegraphics[width=8.0cm]{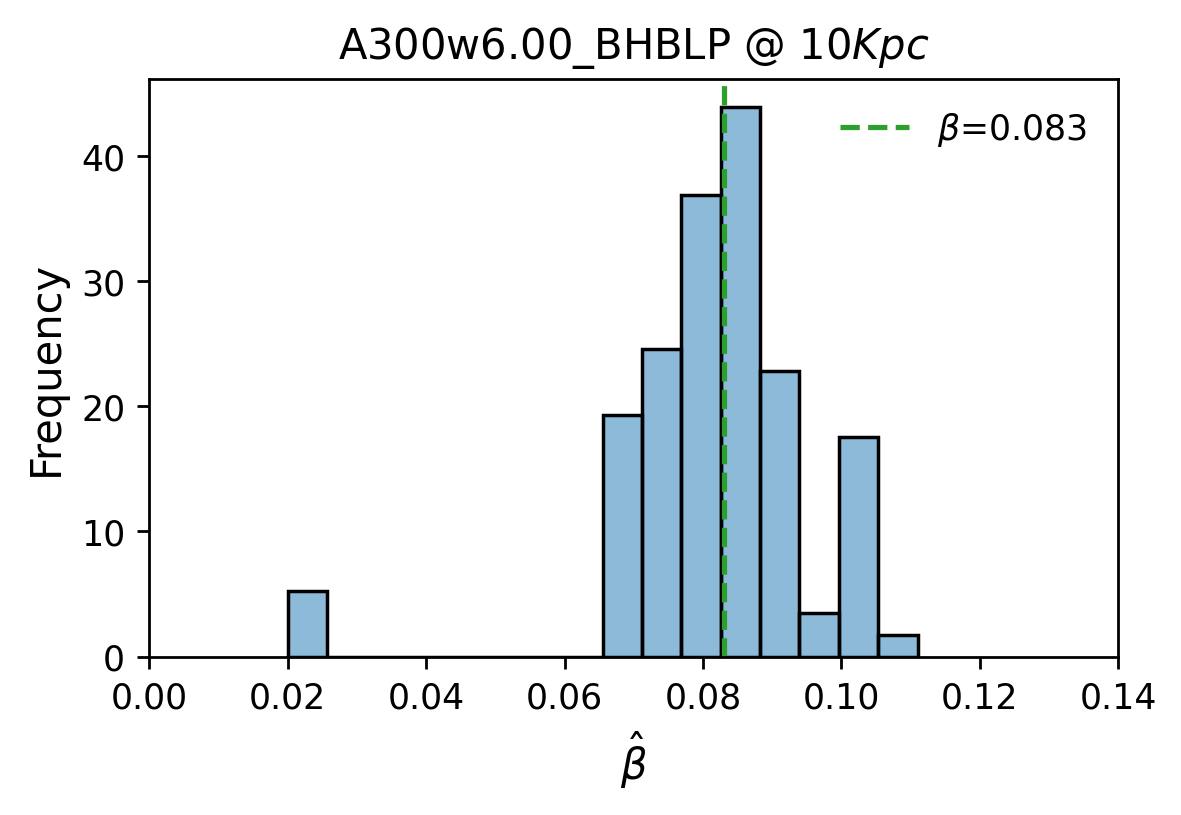}
 \includegraphics[width=8.0cm]{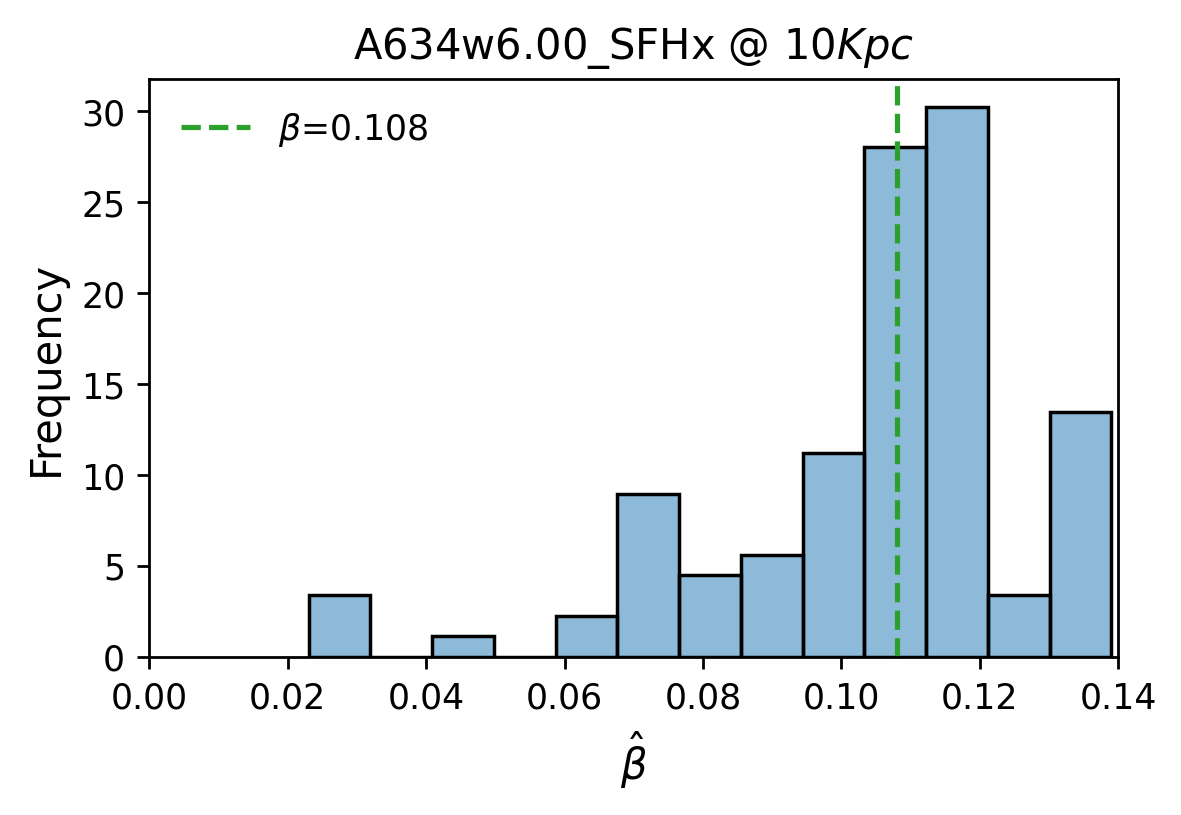}
 \caption{Histogram of the estimated values with MF at distance 10 kpc and optimal orientation. The green dotted line corresponds to the parameter $\beta$ for: top panel the waveform A1268w2.00\_GShenFSU2.1 we have an estimated average value $\hat{\beta}=0.066$ with $\sigma = 1.60\times 10^{-2}$. Middle panel: Correspond to waveform A300w6.00\_BHBLP with an estimated average value $\hat{\beta} =0.082$ and $\sigma = 1.46\times 10^{-2}$. Bottom panel: Finally corresponds to  waveform A634w6.00\_SFHx with an estimated average value $\hat{\beta}=0.105$ and $\sigma = 2.4\times 10^{-2}$.}
\label{Fig:HistogramsMF}
\end{figure}

\begin{figure}[!ht]
 \centering
 \includegraphics[width=8.0cm]{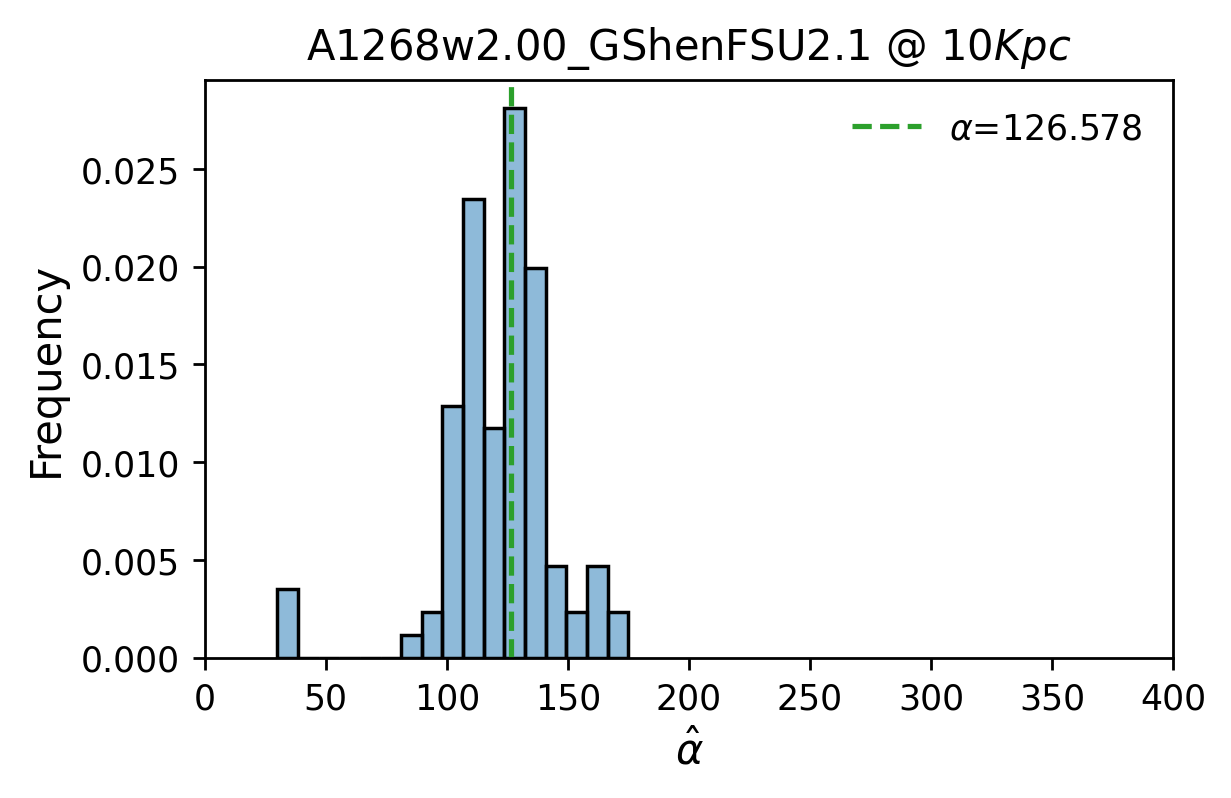}
 \includegraphics[width=8.0cm]{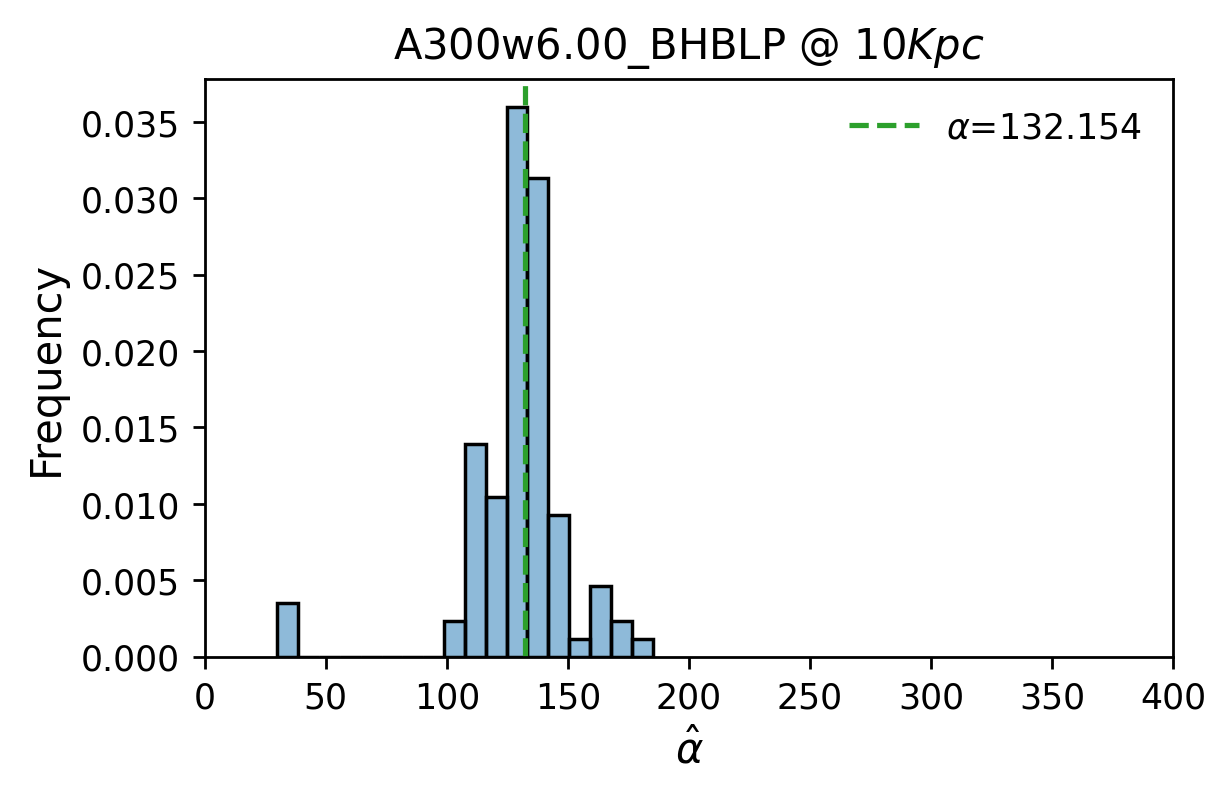}
  \includegraphics[width=8.0cm]{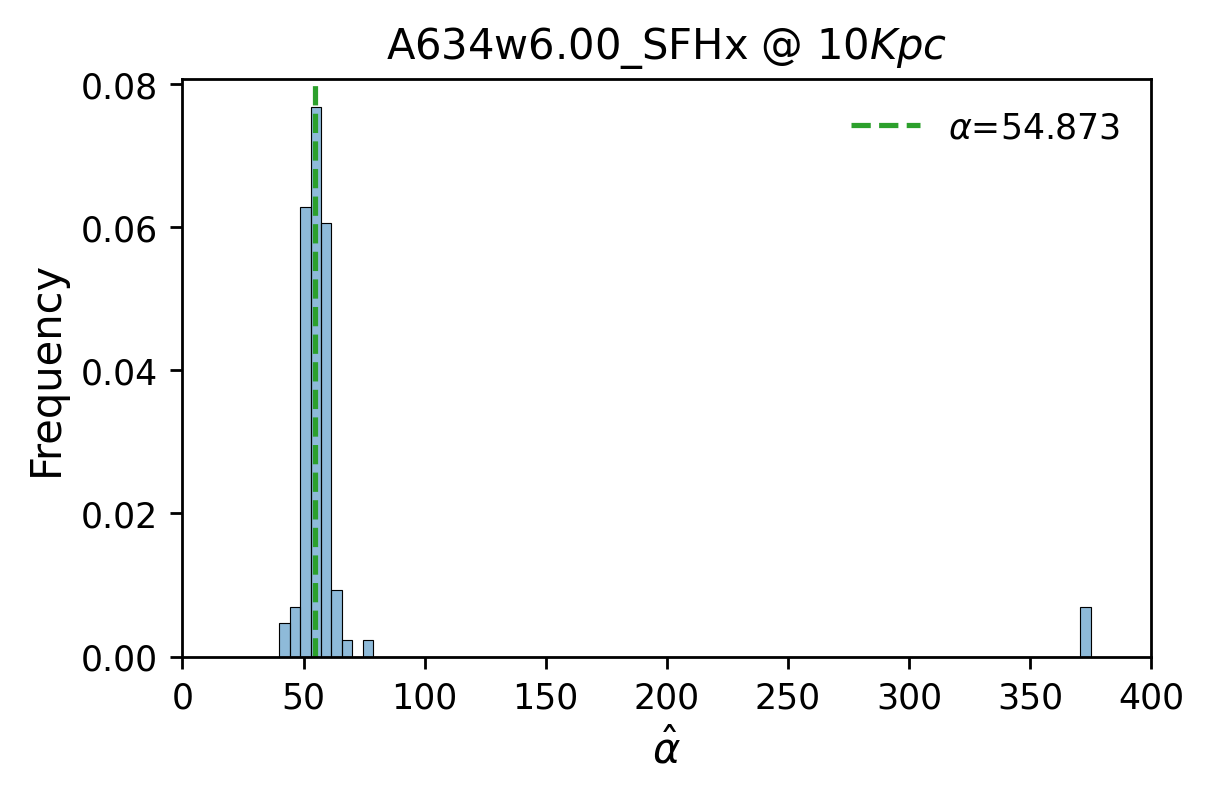}
 \caption{Distribution of the estimate of the parameter $\alpha$ at a distance 10 kpc and optimal orientation. Top panel: for waveform A1268w2.00\_GShenFSU2.1 with $\hat{\alpha}=126.57$ with $\sigma = 22.1$. Middle panel:  waveform A300w6.00\_BHBLP with $\hat{\alpha}=129.50$ with $\sigma=22.7$. Bottom panel: waveform A634w6.00\_SFHx with $\hat{\alpha}=64.70$ with $\sigma=55.1$.}
\label{Fig:MF_alpha}
\end{figure}

On the other hand, we can repeat the MF analysis using the same templates as in the previous case, but this time we will inject the signals in real LIGO noise. Figure \ref{Fig:HistogramsMF} shows the distribution of the estimates of the parameter $\beta$, considering the examples presented above. Also, we use MF to estimated $\alpha$ of different waveforms used in this work. In Figure \ref{Fig:MF_alpha} we show the histogram for the distribution of $\alpha$ estimated  of the three examples mentioned before.

 We find that the average estimated values of A1268w2.00\_GShenFSU2.1 are $\hat{\beta}= 0.069$ with $\hat{\alpha}= 130.5$, for the second example A300w6.00\_BHBLP we have $\hat{\beta}= 0.083$ with $\hat{\alpha}= 132.95$ and finally the estimated values for A634w6.00\_SFHx are $\hat{\beta}= 0.069$ with $\hat{\alpha}= 130.5$. In the three cases, we find that the relative error is of the order of $\sigma \approx 10^{-3}$, an order of magnitude bigger than the case with colored simulated Gaussian noise. This result is relevant to the following analysis of the estimation error calculated from the CRLB.

\begin{figure}[!ht]
 \centering
 \includegraphics[width=8.0cm]{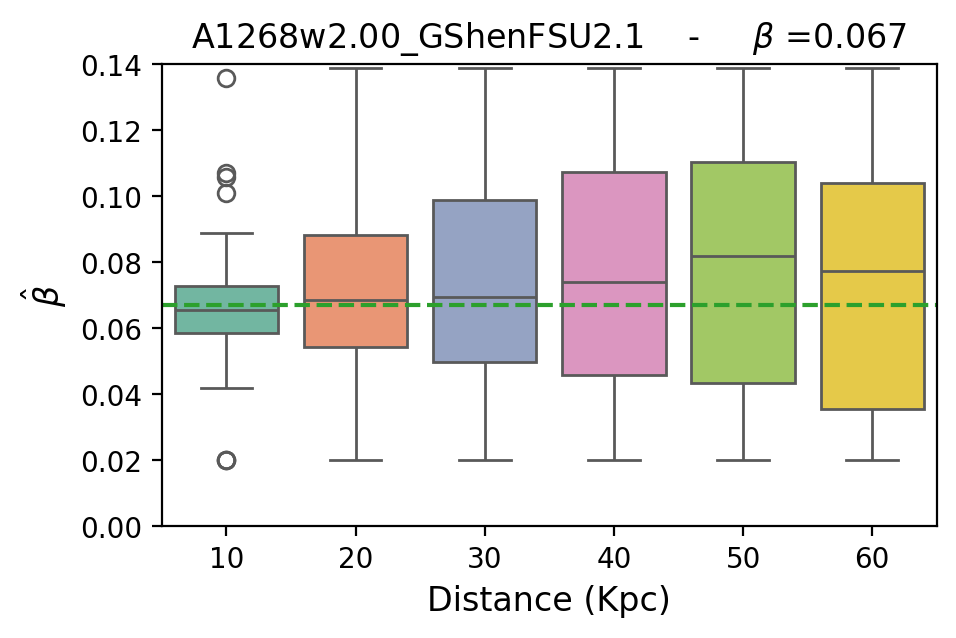}
 \includegraphics[width=8.0cm]{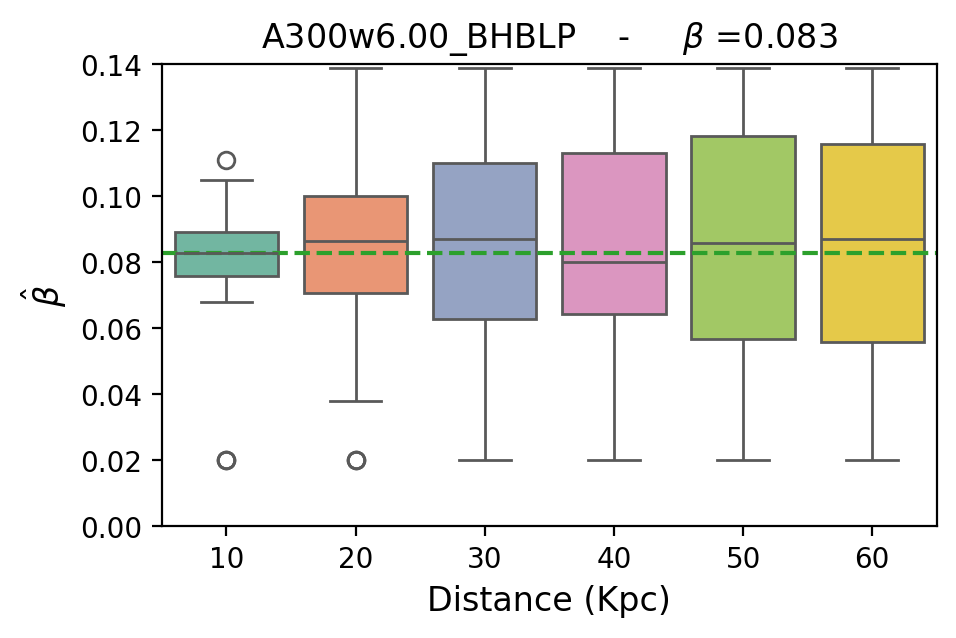}
  \includegraphics[width=8.0cm]{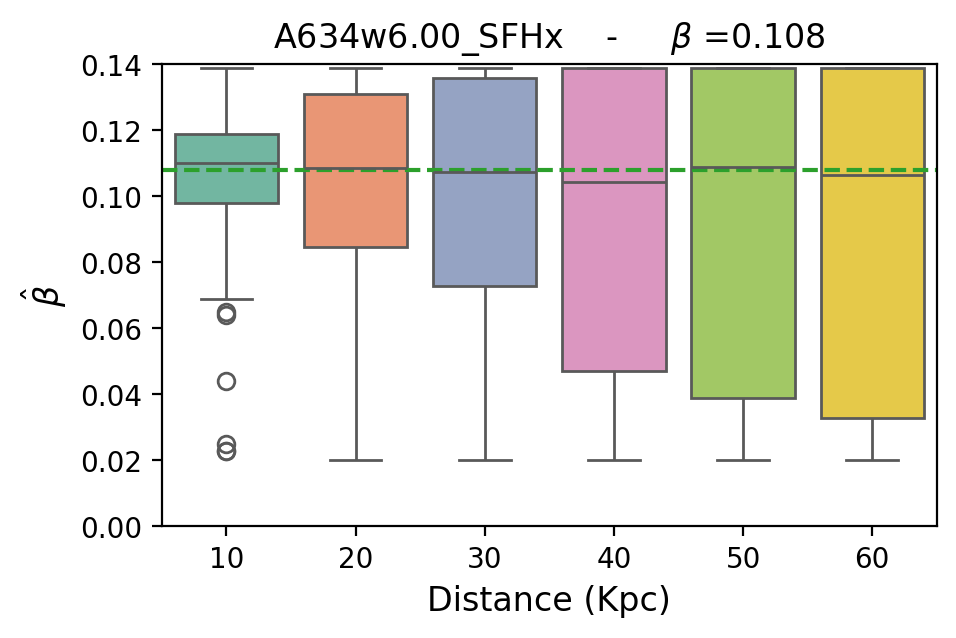}
 \caption{Distribution of the estimate of the parameter $\beta$ at different distances. Top panel for waveform A1268w2.00\_GShenFSU2.1 with $\beta=0.067$. Middle panel we used the waveform A300w6.00\_BHBLP with $\beta=0.083$. Bottom panel is waveform A634w6.00\_SFHx with $\beta=0.108$.}
\label{Fig:MF_distances}
\end{figure}

In the three panels of Figure \ref{Fig:MF_distances}, we show the distribution of the estimate of the parameter $\beta$ at different distances. The green dotted line shows the value of $\hat{\beta}$ obtained from the simulations. The circles represent the outliers, i.e., they are atypical estimates because the combination of parameters can give a waveform different from the expected value. 
%

\section{Theoretical lower bounds on parameter estimation uncertainties} 
\label{Sec:AsymptoticExpansions}

It has been shown \cite{Zanolin_2010, Vitale2011} that the covariance expansions of the MLE can be estimated in terms of the inverse powers of the SNR to obtain the errors in the estimation of parameters. For the first order the variance can be calculated by the CRLB theorem. In \cite{Vitale_2010} it is discussed in detail that the CRLB is inversely proportional to the SNR. Also, in the presence of low SNR or nonlinear estimation, there may be a serious underestimation of the estimation errors and higher orders need to be considered in the expansion.

In order to define the expansion of the variance, let $\textbf{x} =\{x_{1}, ... ,x_{N}\}$ be $N$-dimensional observed data which depends on $P$ unknown parameters $\boldsymbol\vartheta = [\vartheta_1,\vartheta_2,\cdots,\vartheta_P ]^{T}$. 
The observed data are random samples, described by a probability density function (PDF) $p(\textbf{x};\boldsymbol{\vartheta})$.
The goal is to calculate the estimated parameters $\widehat{ \boldsymbol{\vartheta} }$ (hat means estimated) corresponding to $\boldsymbol\vartheta$ based on the observed data $\textbf{x}$. 
We are interested in an estimate $\widehat{ \boldsymbol{\vartheta} }$ that for all parameters is unbiased (i.e., $b(\widehat{\vartheta}_{i}) = \E[\widehat{\vartheta}_{i}] - \vartheta_{i} = 0$ where $i=1,2,\cdots,P$ and $\E[\widehat{\vartheta}_{i}]$ are the expected values of $\widehat{\vartheta}_{i}$) and is as close as possible to the least uncertainty (i.e., $\sigma^2(\widehat{\vartheta}_{i})$ is the minimum achievable variance).
According to the CRLB theorem, the minimum attainable variance of each parameter is $\sigma^2( \widehat{\vartheta_{i}} ) = \textbf{I}_{i j}^{-1} (\boldsymbol{\vartheta}) $, where $\textbf{I}( \boldsymbol{\vartheta} )$ is the $P \times P$ Fisher Information matrix (FIM) given by:
\begin{equation}\label{Equ:FIM}
         \textbf{I}_{ij} (\boldsymbol{\vartheta}) = - \E \left[ \dfrac{\partial^2 \ell (\textbf{x};\boldsymbol{\vartheta})}{\partial \vartheta_{i} \partial \vartheta_{j}} \right],
\end{equation}
and $\ell (\textbf{x};\boldsymbol{\vartheta}) = \ln{(p(\textbf{x};\boldsymbol{\vartheta}))}$ is the log-likelihood function of the unknown parameters given the observed data.

According to the maximum likelihood principle, the MLE $\widehat{ \boldsymbol{\vartheta} }$ of $\boldsymbol{\vartheta}$, which is defined as the value of $\boldsymbol{\vartheta}$ that maximizes $p(\textbf{x};\boldsymbol{\vartheta})$ for a given observed data $\textbf{x}$, is obtained by the stationary point of the log-likelihood as:
\begin{equation}\label{Equ:MLE}
        \ell_{i} (\textbf{x};\boldsymbol{\vartheta}) = \dfrac{\partial \ell (\textbf{x};\boldsymbol{\vartheta})}{\partial \vartheta_{i} } \Bigm|_{ \boldsymbol{\vartheta} = \widehat{\boldsymbol{\vartheta}} } = 0, \; \; \; i=1,2,...,P,
\end{equation}
where $\ell_{i} (\textbf{x};\boldsymbol{\vartheta})$ is the first order partial derivative of the log-likelihood function with respect to the $i$-th parameter.
MLE is important for GW parameter estimation problems because it is asymptotically optimal; that is, for large data records, the MLE has the properties of being unbiased, and if an estimator achieving the CRLB exist, it is the MLE.
In addition, for problems in which observed data are modeled as a signal embedded in noise, as in the case of GW, the MLE is equivalent to the matched filtering performed, for example, for CBC GW signals \cite{Vitale_2010}.
For low SNR, the CRLB might be an under estimation of the error, and using the second order might provide a more accurate lower bound (use the second order of the variance given as inverse expansion series of the SNR). Applying tools for higher-order asymptotic expansions \cite{Bickel1990}, we expand the MLE as a series of inverse orders of SNR \cite{Tso_2016}. It is possible to use the expansion for the variance, considering the expression
\begin{equation}
    \sigma^2 ( \boldsymbol{\vartheta}_{i}) = \sigma^2_{1}(\boldsymbol{\vartheta}_{i}) + \sigma^2_{2}( \boldsymbol{\vartheta}_{i}) + \cdots \, ,
\end{equation}
where $\sigma^2_{1}(\boldsymbol{\vartheta}_{i})$ corresponds to the CRLB, and the explicit expression of $\sigma^2_{2}(\boldsymbol{\vartheta}_{i})$ is given by
%
\begin{eqnarray} \label{Ec:2order}
  \sigma_2^2(\boldsymbol{\vartheta}_{j}) &=& -I^{jm}I^{jn}I^{pq} (v_{nmpq} + 3 \langle h_{nq},h_{pm}\rangle \nonumber \\ 
  &+& 2v_{nmp, q} + v_{mpq,n} ) \nonumber \\
  &+& I^{jm} I^{jn} I^{pz} I^{qt} (v_{npm} v_{qzt} + \frac{5}{2} v_{npq} v_{mzt} \nonumber \\
  &+& 2v_ {qz,n} v_{mtp} + 2v_{qp,z} v_{nmt} + 6 v_{mqp}v_{nt,z} \nonumber \\ &+& v_{pqz}v_{nt,m} + 2v_{mq,z} v_{pt,n} + 2v_{pt,z}v_{mq,n}\nonumber \\
  &+& v_{mz,t}v_{nq,p} )\, ,
\end{eqnarray}
where $I^{ab}$ represents the inverse elements of FIM. For regimes where the MLE is strongly unbiased, and its variance is much larger than the inverse of the FIM, the second order variance also allows us to control the reliability of the first order approximation when evaluating the moments of estimator \cite{Martynov2016}. The physical meaning of the second order variance is related to detecting non gaussianities in the expected histograms of the estimated parameters.
If the error in the estimate induces a Gaussian shape, this indicates that the CRLB is a good approach of the estimate error for a MLE. When the second order is significant, it means that the randomization of the estimate manifests itself as a histogram with a central lobe and sometimes side lobes (this happens for example in direction reconstruction with sonars). Mathematically, and similarly to Taylor expansion approximations, since the derivation of the CRLB contains second order derivatives of the likelihood, it can only capture the curvature of the error distribution. However, the derivation of the second order term involves up to the 4th derivatives, which can also capture the presence of side lobes. 

For problems in which observed data are modeled as a signal embedded in noise, as in the case of GW, the tensors in Equation \eqref{Ec:2order} become:
\begin{eqnarray}
  v_{a,b} &=& -v_{ab}= I_{ab} = \langle h_{a}, h_{b} \rangle \,, \\ 
v_{ab , c} &=& \langle h_{ab} , {h_c}\rangle\,,\nonumber\\
v_{abc} &=& -\langle h_{ab} , h_{c}\rangle -\langle h_{ac} , h_{b}\rangle - \langle h_{bc}, h_{a}\rangle\,,\nonumber\\
v_{ab , cd} &=& \langle h_{ab} , h_{cd}\rangle + \langle h_{a} , h_{b}\rangle \langle h_c , h_d \rangle \,,\nonumber\\
v_{abc , d}&=& \langle h_{abc} , h_d\rangle \,,\nonumber\\
v_{abcd} &=& -\langle h_{ab} , h_{cd}\rangle - \langle h_{ac} , h_{bd}\rangle - \langle h_{ad} , h_{bc}\rangle \nonumber\\
         &&- \langle h_{abc} , h_d\rangle - \langle h_{abd} , h_c\rangle - \langle h_{acd} , h_b\rangle \nonumber\\
         &&- \langle h_{bcd} , h_{a}\rangle \,.\nonumber
\end{eqnarray}
The GW is characterized by the strain $h$ and we can define the derivatives of the frequency domain strain with respect to the parameter space as
\begin{equation}
    h_{a,b,\cdots,P}(f) = \frac{\partial^P h(f)}{\partial \vartheta_a \partial \vartheta_b \cdots \partial \vartheta_P}.
\end{equation}
For GW recorded by a laser interferometer, the noise is additive.
Furthermore unless a glitch is superimposed with the GW, the noise is 
a colored Gaussian process with a variance equal to the LIGO noise PSD (glitches are not expected to affect parameter estimation, but are important for estimating the false alarm rate of search pipelines, which is beyond the scope of this work).  We can express the data observed as
\begin{equation}\label{Equ:SignalPluNoise}
        x(t) = h(t;\boldsymbol{\vartheta}) + n(t),
\end{equation}
where $h(t;\boldsymbol{\vartheta})$ is the signal model and $n(t)$ the noise of the interferometer.

It can be shown that the FIM can be computed in the Fourier domain as
\begin{equation}\label{FIM}
     \textbf{I} (\boldsymbol{\vartheta})_{i j} = \E \left[ \ell_{i} \ell_{j} \right] = \langle h_{i}(f), h_{j}(f) \rangle ,
\end{equation}
where $h_{i}(f) = \partial h(f) / \partial \vartheta_i $ is the derivative with respect to the $i$-th parameter of $h(f) \equiv \int \, \mathrm{\exp} \left(-2\pi i f t \right)\,h(t;\boldsymbol{\vartheta}) \, dt$. 
In Equation \eqref{FIM} we have introduced the mean in the frequency space of two functions $\rm{u({\it f})}$ and $\rm{v({\it f})}$ defined by
\begin{equation} \label{Equ:ScalarProduct}
    \langle {\rm u({\it f}), v({\it f})} \rangle \equiv 4 \, {\rm Re} \int_{{f}_\mathrm{low}}^{{f}_\mathrm{cut}} d{\it f} \, \frac{ {\rm u({\it f})v({\it f})^{*} }}{S_h(f)} \,,
\end{equation}
where $S_h(f)$ is the one-sided PSD of the noise defined as the Fourier transform of the noise auto-correlation function, where the integration range depends on the antenna properties and on the theoretical signal model. 

\subsection{Power Spectral Density}
\label{Sub:PSD}
Calculating the first and second order estimation errors for the rotation parameter $\beta$ (see Equations \eqref{Equ:FIM} and \eqref{Ec:2order}) requires the noise PSD $S_h(f)$. 
Following \cite{Zanolin_2010}, we used the analytical expression of the advanced LIGO noise spectrum, which is interpolated by:
\begin{eqnarray}\label{Ec:PDS}
  S_h(f) = S_0 \left[x^{-4.14} - 5x^{-2} + 111 \frac{1-x^2+x^4/2}{1 + x^2/2} \right],
\end{eqnarray}
where $x = f/f_0$, $f_0 = 215$ Hz and $S_0 = 10^{-49}\,{\rm Hz}^{-1}$. 
This PSD noise is an analytical function defined for $f \geq f_\mathrm{low}$ with $S_h(f) = \infty$ for $f < f_\mathrm{low}$, where $f_\mathrm{low} = 10$ Hz is the detector's lowest frequency cutoff value. 
This analytical expression allows modeling the noise of the LIGO detector, which we can use to calculate the dot products defined by the equation \eqref{Equ:ScalarProduct}. However, in the results presented in this work we will use the real O3 noise data.

Furthermore, we will use PSD data from the Einstein Telescope (ET) sensitivity model from \cite{Hild_2011}. Also, we calculate the PSD using the Amplitude Spectral Density (ASD) of the Cosmic Explorer (CE) detector reported in \cite{Srivastava_2022}.  

\begin{figure}[!ht]
 \centering
 \includegraphics[width=8.0cm]{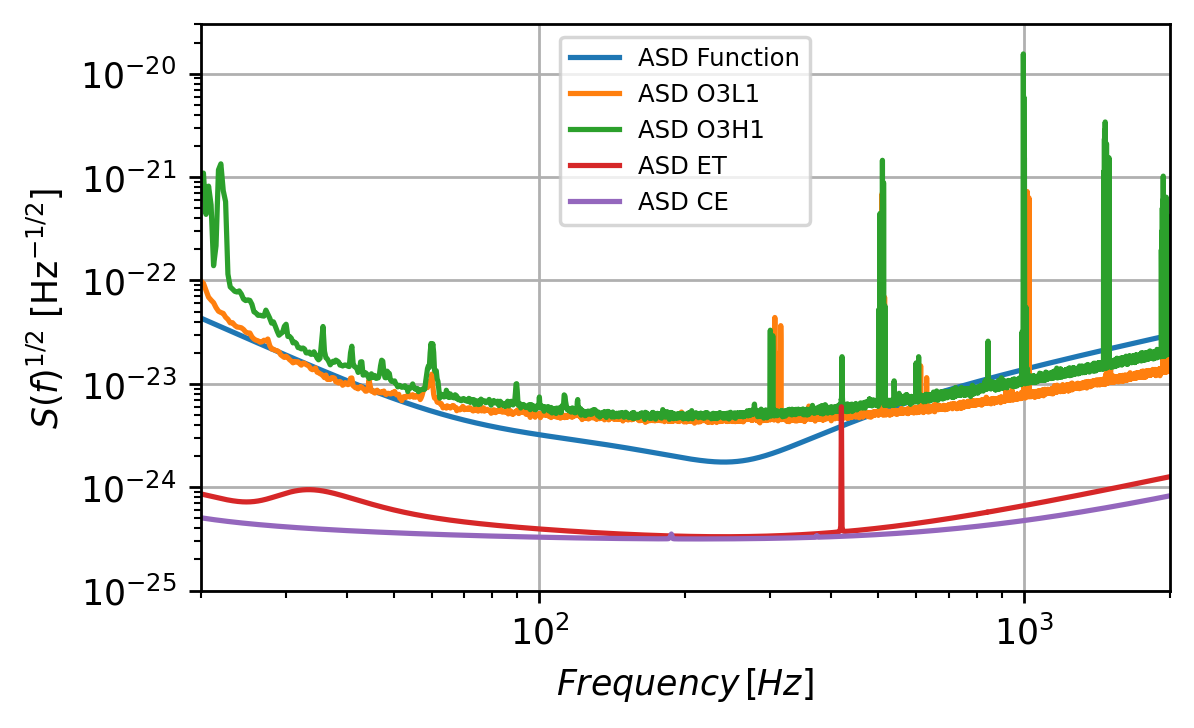}
 \caption{Amplitude Spectral Density. The blue line corresponds to the analytical function $S_h(f)$ while the orange and green lines represent the experimental LIGO noise from the third observing run (O3) in the Livingston and Hanford detectors. The red line represents the Einstein Telescope and the purple line corresponds to the Cosmic Explorer.
 }
\label{Fig:ADS}
\end{figure}

Figure \ref{Fig:ADS} shows this analytical advanced LIGO noise function and the experimental LIGO noise from the third observation run on the Livingston (O3L1) and Hanford (O3H1) detectors \cite{T2000012, Davis_2021}.  In addition, it shows the noise amplitude of ET \cite{Punturo:2010zz} and CE.
%

\subsection{Theoretical estimates of the minimum parameter estimation error }
\label{SubSec:Parameter}

In the CCSNe GW analysis discussed here, we apply the asymptotic expansions for the errors of the MLE, with the aim of calculating the precision of the estimate of the rotation parameter $\beta$. Specifically, we compute the first and second order variance expansions to establish the uncertainties of the $\beta$ estimate. The first order of the variance $\sigma_{1}^2[\beta]$ is given by the inverse of the square root of the Fisher information matrix (FIM), according to the CRLB
\begin{eqnarray} \label{Ec:CRLB}
    \sigma_1^2[\beta] = I_{ij}^{-1} \,.
\end{eqnarray}
Where $I_{ij}$ is the elements of FIM. An important result we can explore is the difference in first-order variance when considering a single parameter (1D) versus considering a three-parameter space (3D). Therefore, we can calculate the Fisher information for one parameter using the equation
\begin{equation} \label{Equ:ScalarProduct_beta}
    I_\beta = 4 \, {\rm Re} \int_{f_{\rm low}}^{f_{\rm cut}} df \, \frac{ {\rm h_{\beta}({\it f}) h_{\beta}^{*}({\it f}) }}{S_h(f)} \,,
\end{equation}
where $h_\beta = \partial h(t)/  \partial \beta$. For the variance $\sigma_1^{-1}[\beta]$ we substitute Equation \eqref{Ec:CRLB}. 
On the other hand, for the 3D case we consider the second element of the diagonal of the inverse FIM $I_{\beta\beta}^{-1}$, which corresponds to 
\begin{equation} \label{Equ:FIM_inverse}
    I_{\beta\beta}^{-1} = \frac{I_{\alpha\alpha}I_{\tau\tau} - I_{\alpha\tau}I_{\tau\alpha}}{\rm{det}(\textbf{I})}.
\end{equation}
We obtain the standard deviation for the first order covariance of the core-bounce phase signal considering a distance of $10$ Kpc. In order to obtain the precision of the variance measurement, we use the relative error $\Delta\, \sigma_1 = \sigma_1/\beta $, which is obtained from the relationship between the variance and the $\beta$ value.

\begin{figure}[!ht]
 \centering
 \includegraphics[width=8.0cm]{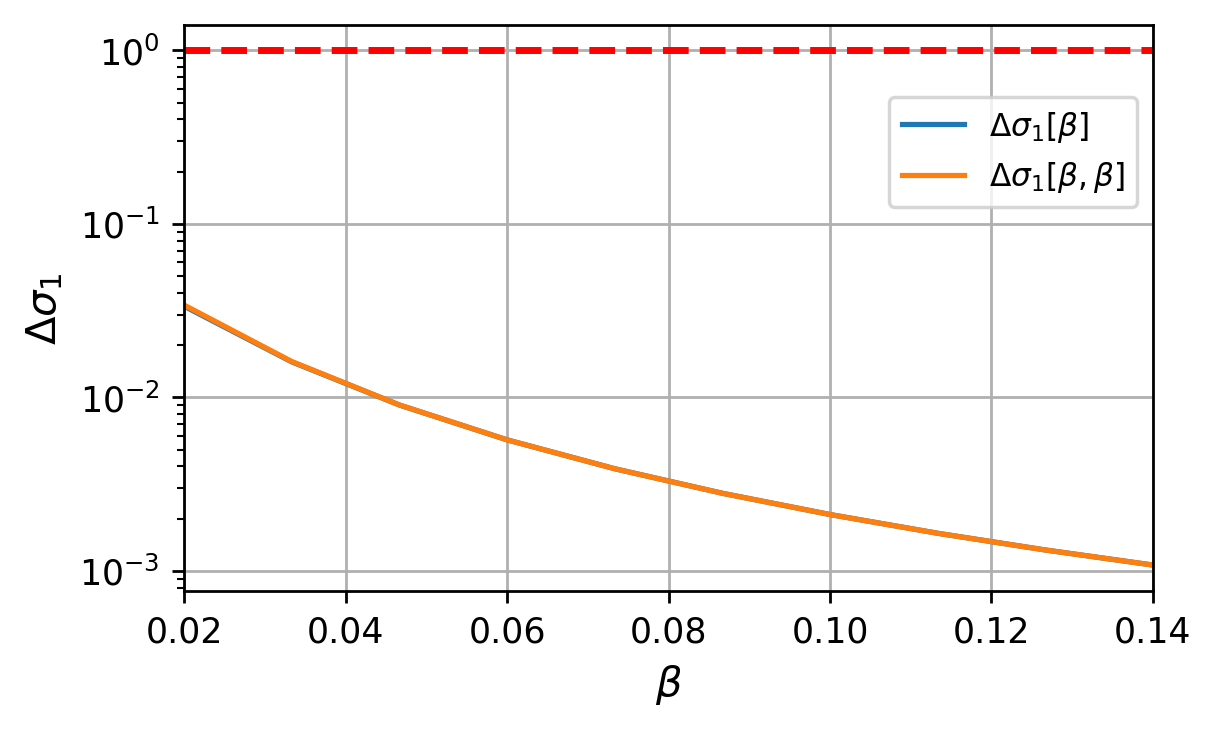}
 \caption{First order variance calculated from CRBL with $\Delta \sigma_1 [\beta]$ for 1D. Also we plot the variance calculated from second element of the diagonal of the invers FIM $\Delta \sigma_1 [\beta,\beta]$ for 3D.}
\label{Fig:comparative}
\end{figure}

\begin{figure}[!ht]
 \centering
 \includegraphics[width=8.0cm]{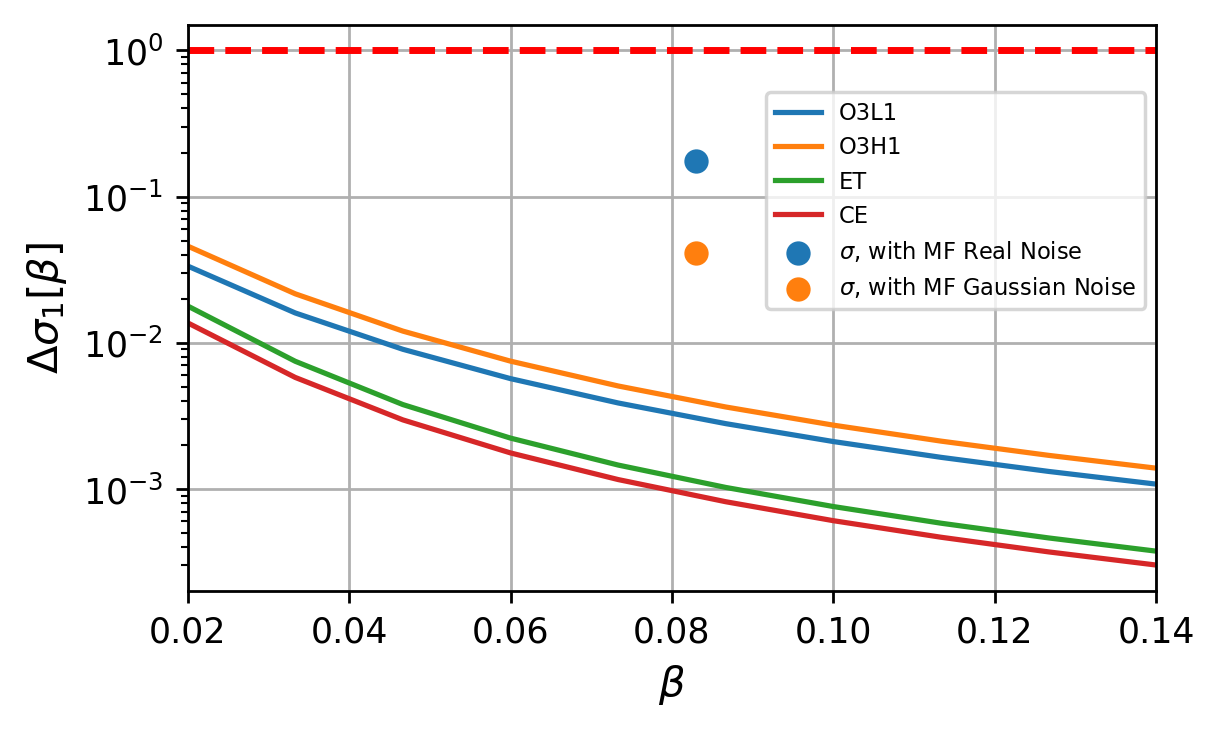}
 \includegraphics[width=8.0cm]{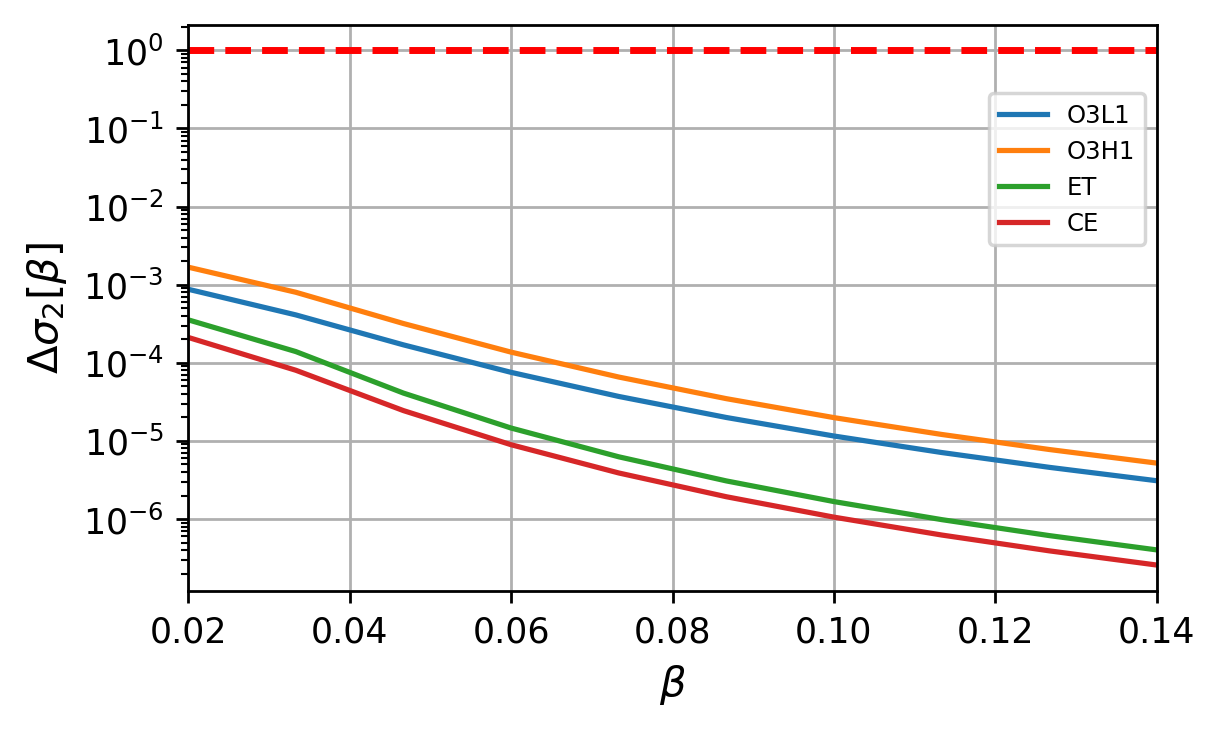}
 \caption{Relative error of variance, as a function of the parameter $\beta$ for signals at a distance of 10 kpc for LIGO (O3L1 and O3H1), ET and CE detector PSD. The horizontal dotted line indicates when the value of the parameter $\beta$ equals the corresponding error, or a relative error equal to one. Top panel: Correspond to relative error of the first order variance $\Delta\, \sigma_1$. The dots correspond to variance of A300w6\_BHBL obtained from MF with real noise (blue) and Gaussian noise (orange). Bottom panel: Correspond to relative error of the second order variance $\Delta\, \sigma_2$. }
\label{Fig:C1}
\end{figure}

According to the results obtained from the 1D parameter space and the 3D parameter space, the difference in first-order variance with a single parameter and a space of three parameters is minimal, as seen in Figure \ref{Fig:comparative}. This indicates that $\beta$ is decoupled from the other two parameters and the 1D results are valid for the 3D case. The variance in this three parameter space is a symmetric matrix, which allows us to find a combination of parameters that can diagonalize it. 
Consequently, we can say that the parameters are independent and allow us to calculate only the elements of the diagonal of the variance.
Now, our analysis focuses on the 3D case again, so we calculated $\sigma_1$ using PSD of LIGO (O3L1 and O3H1), CE and ET detectors, at 10 Kpc from the source in an optimal orientation. The results are shown in the top panel of Figure \ref{Fig:C1}. We see that the relative error decreases as the parameter $\beta$ increases. The red dotted line indicates when the relative error is equal to one. The fact that the theoretical minimum is always smaller than one indicates that in principal should be able to tell that $\beta$ is not zero or in other words detect the present of the rotation in the progenitor. 

\begin{figure}[!ht]
 \centering
 \includegraphics[width=8.0cm]{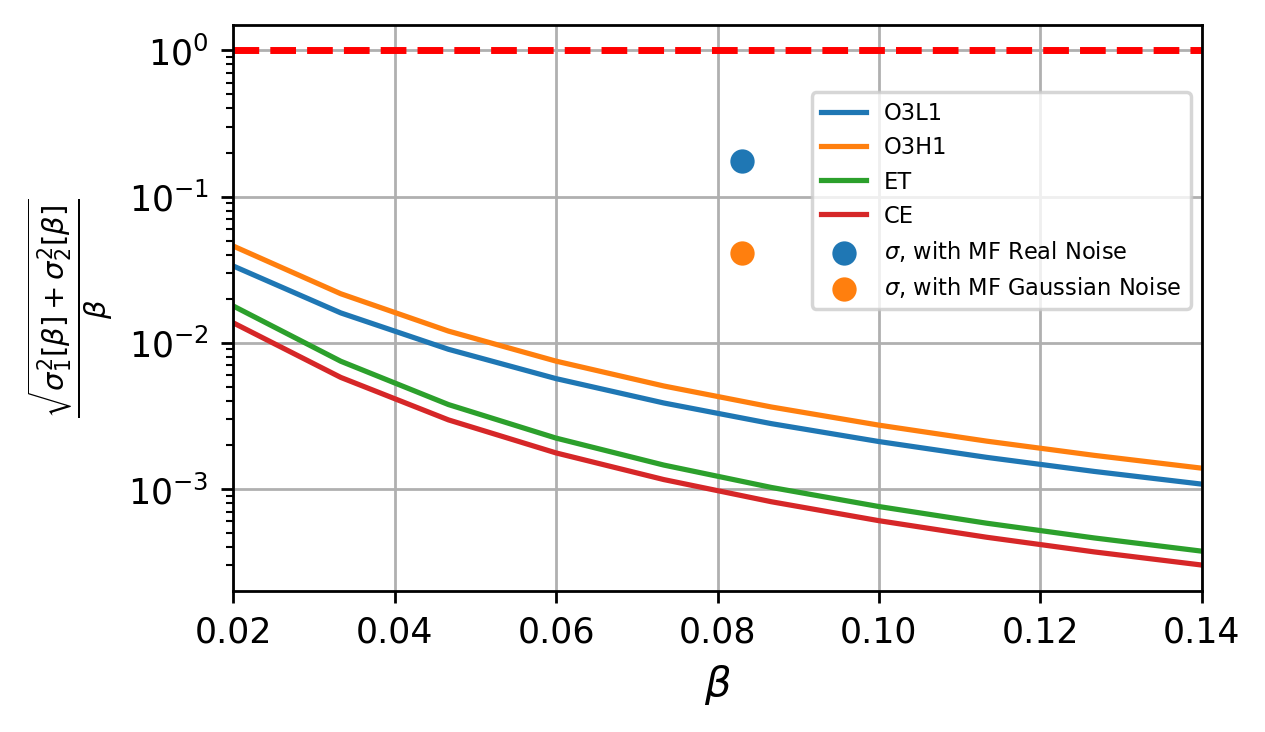}
 \caption{Estimation error of the parameter $\beta$. The horizontal dotted line means that the value of the parameter is equal to the corresponding error. In this plot, we also consider the error to distance 10 kpc for LIGO (O3L1 and O3H1), ET and CE detector PSD. The dots correspond to variance of A300w6\_BHBL obtained from MF with real noise (blue) and Gaussian noise (orange).}
\label{Fig:estimationerror}
\end{figure}

To calculate the second order variance $\sigma_2^2[\beta]$, we use Equation \eqref{Ec:2order}, where it is necessary to consider the entire parameter space. We calculated the relative error for the covariance in second order, using Equation \eqref{Ec:2order} and the definition of the ratio $\Delta\, \sigma_2 = \sigma_2/\beta $. In bottom panel of Figure \ref{Fig:C1} we observe that the relative error decreases as the parameter $\beta$ increases. Comparing two panels of Figure \ref{Fig:C1}, we observe that the relative error of second order covariance can be smaller than the first order by two orders of magnitude, so the second order is not relevant for this analysis. 
So, we obtain the estimation error of the parameter $\beta$, to measure the accuracy of the results obtained by our analytical model,
\begin{equation}
  \frac{\sigma^2}{\beta}=\frac{\sqrt{\sigma_1^2 + \sigma_2^2}}{\beta}.
\end{equation}
We see that the contribution of second order of variance is negligible for the estimation error, so as seen in Figure \ref{Fig:estimationerror} 
This allows us to define a region where the error in the estimation is very small for the case of the RR CCSNe. In the top panel of Figure \ref{Fig:C1} and Figure \ref{Fig:estimationerror} we add as example two dots that correspond at the relative error for A300w6\_BHBLP obtained from MF. The blue dot correspond to standard deviation using real LIGO noise using the PSD for O3L1. And the orange dot is standard deviation for Gaussian noise simulated from the PSD. We estimated $\beta$ with fixed $\alpha$ and $\tau$. Considering the classification used in this article, where $\beta > 0.08$ corresponds to the RR CCSNe, and the estimation error is less than $10^{-1}$ for a distance of $10$ Kpc. The values below the horizontal line in Figure \ref{Fig:estimationerror} define a confidence region for the estimate of the $\beta$ parameter. It is observed that for large values of $\beta$, the relative error is small.
%

\section{Conclusions}
\label{sec:conclusions}

In this paper we discuss the estimation of physical parameters from the Core Bounce component of rapidly rotating core collapse supernova progenitors. 
We propose a model the core-bounce phase of the CCSNe rotational gravitational signals based on three parameters: arrival time, ratio of rotational kinetic energy to potential energy and a phenomenological parameter $\alpha$ related to rotational profiles and EOS.
We quantify the validity of our analytical model using a sample of 126 numerical waveforms from Richers catalogue \cite{Richers_2017}. The 94.4\% average FF between the templates and the numerical simulations is sufficient to estimate the presence of rotation (and $\beta$ different than zero) and the value of $\alpha$ with the uncertainties obtained in this work.
 From MF we conclude that the relative error obtained, for real noise from O3 data at 10 Kpc distance, is of the order $10^{-2}$.
A relative error smaller than 1 indicate the capability to resolve that beta is different than zero and that 
rotation is present in the progenitor. Fig 16 shows a realistic relative error on beta of 0.2. at 10Kpc.
This indicate that the same analysis would have a relative error smaller than 1 at 0.5 Mpc for
third generation interferometers like CE or ET.

We perform 3 types of PE error assessments: MF with real noise, MF with Gaussian noise and estimates of error lower bounds. The differences in the performance between the first and the second are related to non Gaussian features in the noise, while the lower bounds cannot be always be attained. The lower bound are also valuable to understand how errors are expected to change in different regions of the parameter space.
When calculating the FIM considering one parameter ($I_{\beta}$) and comparing it with the $I_{\beta\beta}$ component of the 3-parameter space, Equations \eqref{Equ:ScalarProduct_beta} and \eqref{Equ:FIM_inverse} respectively, the difference between them is minimal, which indicates that $\beta$ is decoupled from the other two parameters and the 1D results are valid for the 3D case. 

In our analysis we consider an optimal orientation and the real noise of the interferometers, mainly the O3L1 data, however we extended the results of the estimation error for the error lower bounds to the Einstein Telescope and Cosmic Explorer. 
 
 We leave the inversion of physical properties from measurements of $\beta$ and $\alpha$ as well as 
the performance for a network of interferometers, versus a single one, for future work.

As an extra note, the amplitude of the wave changes when considering the distance to the source or due to the orientation angle. However, the shape of the wave does not change, i.e. the estimation of $\beta$ is not affected by the degeneracy when considering them independently. There would only be a degeneracy in the estimation if we consider the distance and the angle at the same time, but for nearby CCSNe (within 20 Mpc) the distance is usually known with a small relative error for the goals of this type of analysis.

\begin{figure}[!ht]
 \centering
\includegraphics[width=8.0cm]{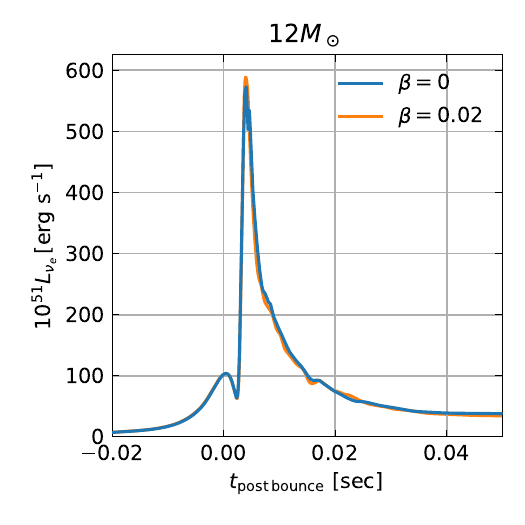}
\caption{Electron type neutrino luminosity (in the fluid frame) versus time for a nonrotating and rotating axisymmetric model from a 12$M_\odot$ progenitor \cite{Sukhbold_2016}.  These results are from models used in \cite{Pajkos_2021}, which utilize a robust neutrino treatment---the so called `M1 scheme' \cite{O_Connor_2018}---and a general relativistic effective potential (GREP) \cite{Marek_2006}. Similar to other models in \cite{Pajkos_2021}, there is slight dependence of $L_{\nu_e}$ on rotation during the bounce, supported by previous work using a different neutrino treatment \cite{Yokozawa_2015}.  Later in the CCSNe, however, rapid rotation can lower neutrino lumonisities.}
\label{Fig:neutrino_burst}
\end{figure}

Estimation of the $\beta$ parameter from the core bounce could be combined with study of other GW characteristics of CCSNe such as the slope of the HFF \cite{casallas2023}. This could allow one to resolve some of the parameter degeneracies involving rotational profile, mass, and EOS. For other CCSNe features like the SASI, a multimessenger approach has proven useful as well. While electron type neutrino luminosities for moderate rotations are only slightly affected (see Figure \ref{Fig:neutrino_burst}), quantifying the impact of rotation on neutrino observations later in the CCSNe is left for future work as well.

%
\begin{acknowledgements}
The authors thank Professors Abdikamalov and Richers for allowing us to use their CCSNe simulation data. We are thanking the national science foundation for the support with NSF2110555. "This material is based upon work supported by NSF's LIGO Laboratory which is a major facility fully funded by the National Science Foundation." This work was supported by CONAHCyT Network Project No. 376127 {\it Sombras, lentes y ondas gravitatorias generadas por objetos compactos astrofísicos}. L.V. acknowledges CONAHCYT scholarship.  C.M. wants to thank PROSNI-UDG and CONAHCYT.  M.A.P. was supported in part by the Sherman Fairchild Foundation, NSF grant PHY-2309231, and OAC-2209655 at Caltech. M.Z. is supported by the National Science Foundation Gravitational Physics Experimental and Data Analysis Program through award PHY-2110555.
\end{acknowledgements}

\appendix

\section{Waveform Selection}
\label{appendix:waveform}

\begin{figure}[!ht]
    \centering
    \includegraphics[width=8.0cm]{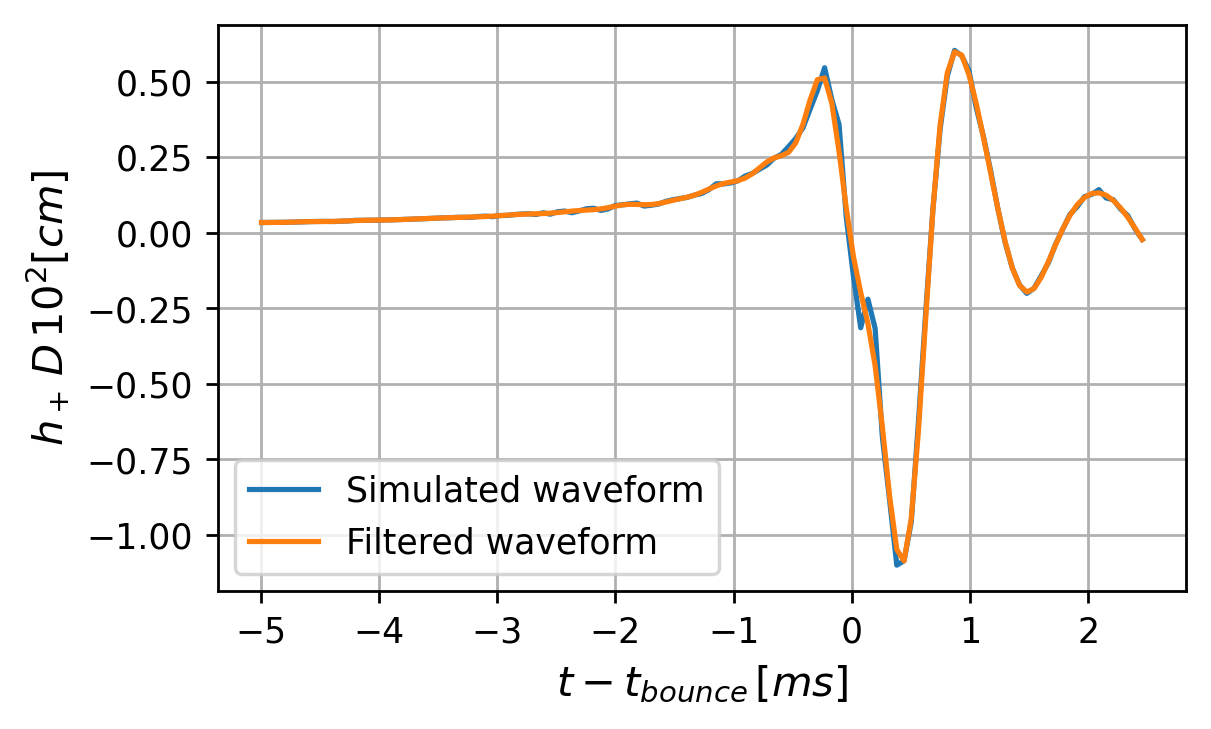}
    \caption{We observe simulated (blue line) and filtered (orange line) waveforms where a numerical artifacts are smoothed out when we use a low pass filter, while preserving the bounce amplitude.} 
    \label{Fig:waveformsfiltered}
\end{figure}

The waveforms chosen for this work were selected from the Richers catalog \cite{Richers_2017}, which has 1824 axisymmetric general relativistic hydrodynamic simulations, covering a parameter space of 98 different rotation profiles and 18 different EOS, sampled at 65535 Hz for a progenitor of 12 $M_\odot$. Commonly used in practice, when calculating the GW strain using the quadrupole formula, an analytic expression is used for the first time derivative of the reduced mass quadrupole moment $\dot{Q}$, and then a finite difference derivative in time is applied to obtain $\ddot{Q}$ \cite{Finn:1990}.  While the simulation may be physically robust, this finite differencing can sometimes lead to minor numerical artifacts in the GW template.  To this end, we apply a filter to preserve the most physical features of each waveform, see Figure \ref{Fig:waveformsfiltered}, without modifying the amplitude and duration of the signal. We discard those that, even when the filter is applied, show numerical artifacts. In addition, we select only EOS that are currently observationally supported by neutron star mass and radius estimates, along with experimental constraints on properties of nuclear matter in each EOS \cite{Richers_2017}: SFHo \cite{steiner_2013}, SFHx \cite{steiner_2013}, LS220 \cite{lattimer_1991}, BHBLP \cite{banik_2014}, HSDD2 \cite{hempel_2010, hempel_2012}, and GShenFSU2.1 \cite{shen_2011}. Taking these numerical and EOS considerations into account, we use 126 simulated waveforms, which cover the values of the parameter space of differential rotation (A) and the maximum initial rotation rate ($\Omega_0$) of the rotation profile \cite{Richers_2017}, as shown in the Table \eqref{Tab:RotationProfiles}.

\begin{table}[!ht]
    \centering
    \begin{tabular}{|c|c|c|}
    \hline
    Name & A[km] & $\Omega_0[{\rm rad\, s}^{-1}]$ \\
    \hline
    A1   & 300   & 3.0 - 11.0 \\ 
    \hline
    A2   & 467   & 3.0 - 6.0  \\
    \hline
    A3   & 634   & 2.0 - 6.0  \\
    \hline
    A4   & 1268  & 1.0 - 5.0  \\
    \hline
    A5   & 10000 & 1.0 - 3.0  \\
    \hline
    \end{tabular}
    \caption{Models used in the analysis, considering the parameters of differential rotation A and maximum rotation speed $\Omega_0$. In total we use 16 rotation profiles and only 6 EOS (SFHo, SFHx, LS220, BHBLP, HSDD2, GShenFSU2.1.).}
    \label{Tab:RotationProfiles}
\end{table}
\section{Frequency of waveforms}
\label{Appendix:freq}

\begin{figure}[!ht]
    \centering
    \includegraphics[width=8.0cm]{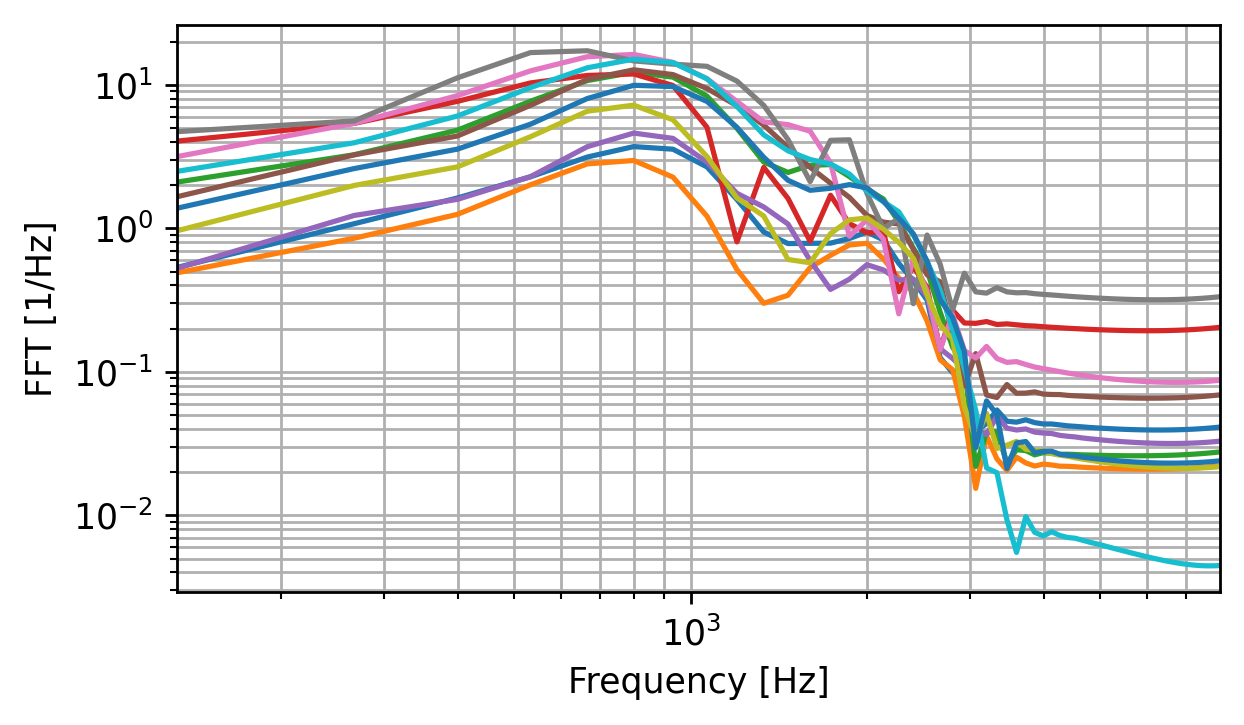}
    \caption{Fourier Transform of a sample waveforms selected for the core bounce phase, assuming a distance of 10 kpc and optimal orientation. We plot only 10 waveforms.}
    \label{Fig:FFT}
\end{figure}

\begin{figure}[!ht]
    \centering
    \includegraphics[width=8.0cm]{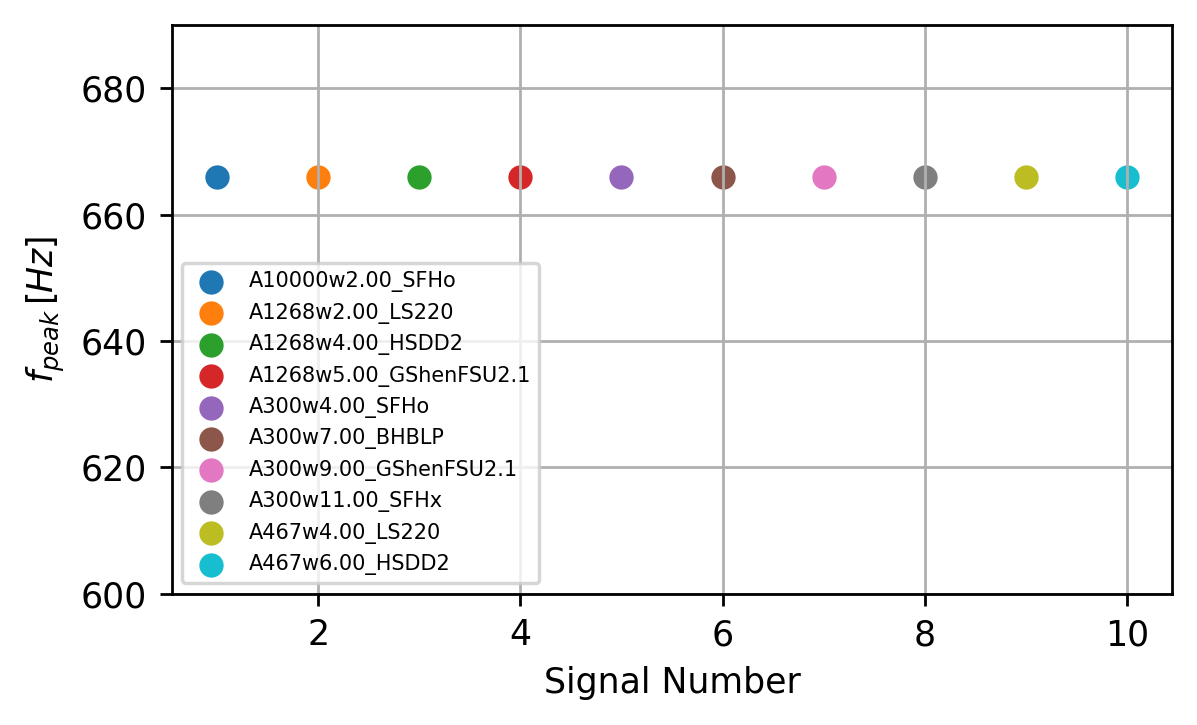}
    \caption{$f_{\rm peak}$ for the a sample of waveforms. We find that $f_{\rm peak}$ value is $666.01626$. }
    \label{Fig:f_peak}
\end{figure}

An important feature we explored in our analysis is the peak frequency $f_{\rm peak}$ of the selected waveforms for this work. Considering only the core bounce phase of the characteristic strain $h(t)$, we switch to the frequency domain using the fast fourier transform (FFT).  In Figure \ref{Fig:FFT} we see only a sample of the 126 signals used. We found that the peak frequency is $f_{\rm peak} = 666.01626$. In the Figure \ref{Fig:f_peak} we see that $f_{peak}$ is constant. 

The small variability of the temporal separation of the 3 peaks of the core bounce phase 
corresponds to the small variability of the peak frequency displayed in Fig 19.

\newpage

\bibliographystyle{unsrt}
\bibliography{References}
\end{document}